\documentclass[useAMS,usenatbib]{mn2e}

\usepackage{graphicx}\usepackage{epsfig}\usepackage{epsf}
\usepackage{amsmath}\usepackage{amssymb}\usepackage{stfloats}
\usepackage{soul}
\usepackage[usenames]{color}
\definecolor{green}{rgb}{0.3,0.7,0.}
\title[Gaia LBVs]{On the Gaia DR2 distances for Galactic Luminous Blue Variables}

\author[Smith et al.]{Nathan Smith$^{1}$\thanks{E-mail:
    nathans@as.arizona.edu}, Mojgan Aghakhanloo$^2$, Jeremiah
  W. Murphy$^2$, Maria R. \newauthor
  Drout$^{3,4}$, Keivan G. Stassun$^{5,6}$,  Jose H. Groh$^7$ \\ $^1$Steward Observatory, 
  University of Arizona, 933 N. Cherry Ave., Tucson, AZ 85721, USA \\ $^2$1Department of 
  Physics, Florida State University, 77 Chieftan Way, Tallahassee, FL 32306, USA \\ $^3$The 
  Observatories of the Carnegie Institution for Science, 813 Santa Barbara St., Pasadena, 
  CA 91101, USA \\ $^4$Hubble and Carnegie-Dunlap Fellow \\ $^5$Vanderbilt University,
  Department of Physics \& Astronomy, 6301 Stevenson Center Lane,
  Nashville, TN 37235, USA \\ $^6$Fisk University, Department of
  Physics, 1000 17th Avenue N., Nashville, TN 37208, USA \\
   $^7$School of Physics, Trinity College Dublin, the University of Dublin, Dublin, Ireland}

\begin{document}

\pagerange{\pageref{firstpage}--\pageref{lastpage}} \pubyear{2012}
\maketitle
\label{firstpage}

\begin{abstract}
We examine parallaxes and distances for Galactic luminous blue
variables (LBVs) in the {\it Gaia} second data release (DR2).  The sample
includes 11 LBVs and 14 LBV candidates.  For about half of the sample,
DR2 distances are either similar to commonly adopted literature values, 
or the DR2 values have large uncertainties.
For the rest, reliable DR2 distances differ significantly
from values in the literature, and in most cases the {\it Gaia} DR2
distance is smaller.  Two key results are that the S~Doradus
instability strip may not be as clearly defined as previously thought,
and that there exists a population of LBVs at relatively low
luminosities.  LBVs seem to occupy a wide swath from the end of the main 
sequence at the blue edge to $\sim$8000~K at the red side, with a spread 
in luminosity reaching as low as log($L$/$L_{\odot}$)$\approx$4.5.  The lower-luminosity 
group corresponds to effective single-star initial masses of 10-25~$M_{\odot}$, and
includes objects that have been considered as confirmed LBVs.  We discuss 
implications for LBVs including (1)~their instability and origin in binary evolution, 
(2)~connections to some supernova (SN) impostors such as the class of SN~2008S-like objects, 
and (3)~LBVs that may be progenitors of SNe with dense circumstellar material 
across a wide initial mass range.  Although some of the {\it Gaia} DR2 distances for 
LBVs have large uncertainty, this represents the most direct and consistent set of 
Galactic LBV distance estimates available in the literature.
\end{abstract}


\section{INTRODUCTION}\label{sec:intro}

Luminous blue variables (LBVs) are the brightest blue irregular
variable stars in any large star-forming galaxy.  They can achieve the
highest mass-loss rates of any known types of stars, and they exhibit
a wide diversity of irregular and eruptive variability
\citep{conti84,hd94,vg01,clark05,smith04,so06,smith11,smith14,vdm12}.  Yet
despite decades of study, the physical mechanism that causes their
variability remains unknown.  An important corollary is that there are
stars that occupy the same parts of the Hertzsprung-Russell (HR)
diagram that are not (currently) susceptible to the same instability. 
The degree to which a star in this region of the HR diagram is unstable 
may depend on its initial mass, its age, its history of mass loss 
(and hence, its metallicity), and past binary interaction.

For any class of stars, distances and true bolometric luminosities are
important for understanding their physical nature.  This is
particularly true for LBVs, since their defining instability, mass loss, and
evolutionary state are thought to be a consequence of their high
luminosity \citep{conti84,hd94,lf88,uf98,owocki04}.  Precise distances 
are critical for inferring whether a
star is in close proximity to the classical Eddington limit based on
its position on the HR diagram compared to stellar evolution
model tracks.  Many LBVs seem to skirt the observed upper luminosity boundary
on the HR diagram, the Humphreys-Davidson (HD) limit,
oscillating between their hot quiescent states and cooler eruptive
states when they cross that observational limit \citep{hd94}.  Based on distances,
luminosities, and temperatures estimated for a few LBVs in the
Milky Way and nearby galaxies, \citet{wolf89} proposed that LBVs in
their hot states reside along the S~Doradus instability strip on the
HR Diagram.  This zone in the HR Diagram is thought to be an important
clue to their instability, perhaps related to the Eddington limit
modified by opacity \citep{lf88,uf98} and rotation \citep{groh09b,langer99}.

Extragalactic LBVs in the Large or Small Magellanic Clouds (LMC/SMC)
and in the nearby spirals M31/M33 have reliable distance estimates, and
hence, fairly reliable estimates of their bolometric luminosity if
detailed quantitative analysis has been used to estimate their
bolometric corrections.  In distant environments, however, we may be
missing faint LBVs, if they exist, either due to selection effects 
(low amplitude variability, for example) or
because they have not received as much observational attention as the most
luminous stars. Noticing that a star is actually an LBV is often the
result of detailed analysis and long-term monitoring; a typical
S~Doradus cycle of an LBV may last a decade.  Moreover, difficulties
associated with contaminating light from neighboring stars become more
problematic for extragalactic LBVs.

Distances to LBVs cannot be determined solely by detailed
spectroscopic analysis of an individual star, because the relationship
between the spectrum and absolute luminosity is ambiguous 
\citep{najarro97,Hillier+98,groh09}.  The
problem is that the emission-line spectra of LBVs can be dominated by
wind emission, which depends on density and ionization of the outflow,
not the absolute luminosity of the star or its surface gravity.  Stars
with dense winds can have very similar spectra across a wide range of
luminosity. For instance, \citet{groh13} have shown that spectral
models of an evolved 20-25 $M_{\odot}$ star that is moving blueward
after experiencing high mass loss in the RSG phase can have a spectrum that closely resembles a
vastly more luminous classical LBV like AG Car.  As such, other types
of evolved stars at lower mass that have effective temperatures similar 
to LBVs and dense winds or disks like a
B[e] star, post-AGB star, or various types of interacting binaries can
have similar emission-line spectra that may masquerade as LBVs.  These
can be mistaken for more luminous LBVs if assumed to be at a distance
that is too far, and vice versa.  We will see below that this is
indeed the case for a few objects that have been considered LBVs or
LBV candidates (a candidate, referred to as cLBV here, is a supergiant star that resembles
an LBV spectroscopically or has a shell nebula, but has not yet been seen to exhibit
the tell-tale variability of either S~Doradus cycles or a giant eruption). 
In addition to the luminosity, other stellar parameters derived from 
spectroscopic analysis also depend on the assumed distance $d$. The stellar 
radius depends on $d$ (relevant for e.g. binary interaction), and mass-loss 
rates scale as $d^{1.5}$.  This influences our interpretation of the 
mass-loss history, fate, circumstellar material properties in interacting 
supernovae (SNe), etc.  Other properties like the effective temperature 
$T_{\rm eff}$ and the terminal wind speed $v_{\infty}$ have a negligible 
dependence on the distance (see \citealt{groh09}).

Of course, star clusters have been a useful tool for estimating
distances and ages for many classes of stars.  A significant
impediment to determining LBV distances by this method, however, is
that many LBVs are not associated with clusters, counter to
expectations for the origin of LBVs in single-star evolution (e.g.,
\citealt{ln02}).  \citet{st15} showed that LBVs are isolated from
clusters of O-type stars in general, whereas the few that are in
clusters seem to be too young (or overluminous) for their environment.
\citet{mojgan17} showed that a passive cluster dispersal model can reproduce
the observed statistical spatial distribution of O-type stars on the
sky, but cannot explain LBVs if they are the descendants of those O-type
stars as expected in single-star evolutionary models.  This may
indicate instead that LBVs are massive blue stragglers and that their
apparent isolation arises either because they received a kick when
their companion star exploded, or because they have been rejuvenated
by mass accretion or mergers in binary evolution
\citep{kg85,jsg89,st15,smith16,mojgan17}.  This might make their
surrounding stellar populations look much older than we would naively
expect from an LBV's position on the HR Diagram.  If LBVs really are
the product of binary interaction, this has important implications for
the origin of their instability.  Stars that have arrived at the same point on the HR diagram through different evolutionary trajectories (single star, mass donor, mass gainer, merger) might help explain why some stars in this part of the HR diagram suffer from eruptive instability and some do not.  Indeed, \citet{justham14} have
discussed the hypothesis that LBVs result from stellar mergers for
theoretical reasons unrelated to their environments ---  in particular, 
that they might be viable SN progenitors.

These inferences about age and environments of LBVs were based on
stars in the LMC, where the distance is reliable.  Such environmental
comparisons are harder in the Milky Way because of distance
ambiguities and extinction.  For this reason, \citet{st15} did not
undertake a quantitative assessment of LBV isolation for Milky Way
LBVs (although they did note anecdotal evidence that Galactic LBVs do appear
remarkably isolated as well).  A Milky Way star seen near other O-type
stars on the sky might be at a different distance but seen nearby in
projection, or alternatively, a lack of O stars in the vicinity might
be a selection effect (LBVs are very bright at visual wavelengths, but
hotter and visually fainter main-sequence O-type stars might be dim
and possibly undetected because of extinction in the Galactic plane).
These complications make it difficult to know if the statistical
environments of LBVs in the LMC also apply in the Milky Way, where the
metallicity sensitivity of the LBV instability might differ.
Similarly, the lack of reliable distances for Milky Way LBVs has
hampered our understanding of their true physical parameters,
especially their bolometric luminosities.  Since most LBVs are not
associated with young clusters of O-type stars, many of them have very
uncertain distances in the literature (see Section 2.2), and similarly, highly 
uncertain ages and initial masses.

Of the dozen or so LBVs in the Milky Way \citep{clark05}, only a few
are seen to be associated with massive young clusters or associations.   
One is $\eta$ Car, arguably the most massive and luminous star in the
Milky Way.  The others are FMM~362 and the Pistol Star, both
apparently associated with the Quintuplet Cluster in the Galactic
Center and visually obscured (and therefore not relevant to the statistical 
assessments of LBV isolation, since we do not have a meaningful sample of 
visually obscured LBVs in the field).  The other is W243 in the massive 
young cluster Westerlund 1 (Wd1).  (Wra~751 is seen amid a cluster too, but not a 
massive young cluster.)  
Here we compile the distances for unobscured Galactic LBVs that
have measured parallax values included in the {\it Gaia} second data release 
(DR2), and we comment on the revised HR diagram for LBVs.

\begin{center}
\begin{table*}\begin{minipage}{7.0in}%
    \caption{Previous Literature Distances for Galactic LBVs and candidate LBVs (in parentheses)}\scriptsize
\begin{tabular}{l|cc|ccc} 
\hline
 & \multicolumn{2}{|c|}{Baseline Distance} & \multicolumn{3}{|c|}{Original Studies}  \\  
 \hline
Name   &$d_{\rm lit}$ (kpc) &Ref.\footnote{N12: \citet{naze12}; G12: \citet{Gvaramadze2012}; K16: \citet{Kniazev2016}; S7: \citet{smith07}.} & $d_{\rm lit}$ (kpc)  & Ref. & Technique \\ \hline
HR~Car        &5.20 &N12 &5.4$\pm$0.4 & \cite{Hutsemekers1991} & Kinematics  \\
    &               &    &5$\pm$1 & \cite{vanGenderen1991} & Reddening-Distance Relationship \\
AG~Car        &6.00 &N12 &$>$5 & \cite{Humphreys1989} & Reddening-Distance Relationship  \\
    &               &    &6.4$-$6.9 & \cite{Humphreys1989} & Kinematics  \\
    &               &    &6$\pm$1 & \cite{Hoekzema1992} & Reddening-Distance Relationship  \\
Wra~751       &6.00 &N12 &$\sim$6 & \cite{pasquali06} & Kinematics (private communication)  \\
    &               &    &$>$5 & \cite{Hu1990} & Reddening-Distance Relationship  \\
    &               &    &$>$4$-$5 & \cite{vanGenderen1992} & Reddening-Distance Relationship  \\
Wd1 W243      &5.00 &N12 &$<$ 5 & \cite{clark05b} & Cluster Membership - YHG Luminosity \\
    &               &    &$>$2 & \cite{clark05b} & Cluster Membership - WR Stars \\
    &               &    &5$^{+0.5}_{-1.0}$ & \cite{Crowther2006} & Cluster Membership - WR Stars  \\
    &               &    &3.55$\pm$0.17 & \cite{Brandner2008} & Cluster Membership - MS Fitting  \\
HD 160529     &2.50 &N12 &$\sim$2.5 & \cite{Sterken1991} & Comparison with R110  \\
HD 168607     &2.20 &N12 &DM$=$11.86 & \cite{Humphreys1978} & Cluster Membership - MS fitting  \\
    &               &    &2.2$\pm$0.2 & \cite{Chini1980} & Cluster Membership - MS Fitting  \\
P Cyg         &1.70 &N12 &1.7$\pm$0.1 & \cite{najarro97} & Spectral Modeling  \\
MWC 930       &3.50 &N12 &3$-$4 & \cite{Miroshnichenko2005} & Associated with the Norma Spiral Arm  \\
    &               &    &  & & Kinematics, Extinction arguments \\
G24.73+0.69   &5.20 &N12 &$<$5.2 & \cite{clark2003} & Reddening-Distance Relationship  \\
%What about G26.47+0.02?
WS~1          &11.0 &G12 &$\sim$11 & \cite{Gvaramadze2012} & Assume Luminosity \\
MN48          &5.00 &K16 &3$-$5 & \cite{Kniazev2016} & Assume association with spiral arm  \\
    &               &    &      &  & Luminosity, Kinematics arguments  \\
\hline
(HD80077)     &3.00 &N12 &2.8$\pm$0.4 & \cite{Steemers1986} & Cluster Membership - MS Fitting  \\
    &               &    &$\sim$3.2 & \cite{Moffat1977} & Cluster Membership - MS Fitting  \\
(SBW1)        &7.00 &S7  &$\sim$7 & \cite{sbw07} & Luminosity Class, Kinematics Arguments \\
(Hen 3-519)   &8.00 &N12 &$>$6 ($\sim$8) & \cite{Davidson1993} & Reddening-Distance Relationship \\
(Sher 25)     &6.30 &N12 &6.3$\pm$0.6 & \cite{Pandey2000} & Cluster Membership - MS fitting \\
    &               &    &6.1$\pm$0.6 & \cite{dePree1999} & Kinematics  \\
($\zeta^1$Sco)&2.00 &N12 &2.0$\pm$0.2 & \cite{Baume1999} & Cluster Membership - MS fitting  \\
(HD 326823)   &2.00 &N12 &$>$2 & \cite{McGregor1988} &  Location in direction of galactic center \\
    &               &    &1.98 & \cite{Kozok1985} & Luminosity Class, Color-Luminosity Relation \\
(WRAY 17-96)  &4.50 &N12 &$<$4.9 & \cite{egan02} & Reddening-Distance Relationship  \\
(HD 316285)   &1.90 &N12 &$\sim$2 & \cite{Hillier+98} & Assumed, Reddening Arguments \\
    &               &    &$\sim$3.4 & \cite{vanderveed1994} & Assumed, Reddening Arguments  \\
(HD 168625)   &2.20 &N12 &DM$=$11.86 & \cite{Humphreys1978} & Cluster Membership - MS fitting  \\
    &               &    &2.2$\pm$0.2 & \cite{Chini1980} & Cluster Membership - MS Fitting \\
(AS 314)      &8.00 &N12 &$\sim$10 & \cite{miro00} & Assume Luminosity \\
(MWC 314)     &3.00 &N12 &3.0$\pm$0.2 & \cite{Miroshnichenko1998} &  Kinematics \\
(W51 LS1)     &6.00 &N12 &8.5 $\pm$2.5 & \cite{Schneps1981} & maser proper motion - membership \\
    &               &    &6.1$\pm$1.3 & \cite{Iamai2002} & maser proper motion - membership \\
    &               &    &$\sim$5.5 & \cite{Russeil2003} & kinematics  \\
    &               &    &$<$5.8 & \cite{Barbosa2008} & Radio Luminosity/Lyman Continuum Photons \\
    &               &    &2.0$\pm$0.3 & \cite{Figueredo2008} & spectroscopic parallax  \\
    &               &    &5.1$^{+2.9}_{-1.4}$ & \cite{Xu2009} & trigonometric parallax + maser proper motion \\
    &               &    &5.41$^{+0.31}_{-0.28}$ & \cite{Sato2010} & trigonometric parallax + maser proper motion \\
(G79.29+0.46) &2.00 &N12 &1$-$5, assume 2 & \cite{Higgs1994} & Reddening-Distance, Kinematics  \\
    &               &    &1$-$4, assume 2 & \cite{Voors2000} & Reddening-Distance, Kinematics \\
(CYG OB2 12)  &1.70 &N12 &1.7$\pm$0.2 & \cite{Torres-Dodgen1991} & Cluster Membership - MS fitting \\
    &               &    &DM=11.4$\pm$0.1 & \cite{Massey1991} & Cluster Membership - MS Fitting \\
\hline\end{tabular}
\label{tab:tab4}\end{minipage}
\end{table*}
\end{center}

\begin{center}
\begin{table*}\begin{minipage}{6.5in}%
    \caption{Parameters from the {\it Gaia} DR2 and Bailer-Jones catalogs.  }\scriptsize
\begin{tabular}{lccccccccccc} 
\hline
Name          &
Gaia DR2 id & 
$\varpi$ (mas)  &
$\sigma_{\varpi}$ (mas) &
$N$ &
RUWE &
$\epsilon$ (mas) & 
$D$ & 
$\ell$ (kpc) &
$d_{\text{BJ}}$ & 
$d_{\text{BJ,low}}$ (kpc) & 
$d_{\text{BJ,high}}$\footnote{The following are from the {\it Gaia} DR2 catalog: $\varpi$ is the parallax, $\sigma_{\varpi}$ is the expected uncertainty in the parallax, $N$ is the number of visibility periods, RUWE is the re-normalized goodness of fit ($\sqrt{\chi^2/(N-5)}$), $\epsilon$ is the excess astrometric noise, $D$ is the significance of the excess astrometric noise.  The following are from the Bailer-Jones catalog: $\ell$ is the length scale of field stars in the direction of the LBV or candidate, $d_{BJ}$ is the most likely Bailer-Jones distance, $d_{BJ,low}$ and $d_{BJ,hihg}$ give the highest 68\% density interval (HDI).}
\\
\hline
HR~Car & 5255045082580350080 & 0.171 & 0.033 & 18 & 1.04 & 0.000 &   0.0 & 1.66 & 4.89 & 4.20 & 5.82 \\
AG~Car & 5338220285385672064 & 0.153 & 0.029 & 17 & 0.88 & 0.000 &   0.0 & 1.60 & 5.32 & 4.59 & 6.29 \\
Wra~751 & 5337309477433273728 & 0.169 & 0.044 & 19 & 1.02 & 0.149 &   8.3 & 1.57 & 4.82 & 3.97 & 6.06 \\
Wd1 W243 & 5940105830990286208 & 0.979 & 0.165 & 11 & 1.11 & 0.582 & 120.1 & 1.38 & 1.03 & 0.86 & 1.27 \\
HD 160529 & 4053887521876855808 & 0.438 & 0.057 & 10 & 0.94 & 0.000 &   0.0 & 2.15 & 2.18 & 1.92 & 2.50 \\
HD 168607 & 4097791502146559872 & 0.644 & 0.060 & 10 & 1.02 & 0.000 &   0.0 & 1.52 & 1.50 & 1.37 & 1.65 \\
P Cyg & 2061242908036996352 & 0.736 & 0.180 & 17 & 1.05 & 1.085 & 437.5 & 1.16 & 1.37 & 1.06 & 1.93 \\
MWC 930 & 4159973866869462784 & -0.162 & 0.094 & 11 & 0.74 & 0.217 &  10.0 & 1.36 & 7.81 & 5.78 & 10.64 \\
G24.73+0.69 & 4255908794692238848 & -0.329 & 0.223 & 10 & 1.03 & 1.302 & 233.6 & 1.44 & 5.44 & 3.58 & 8.25 \\
WS~1 & 4661784273646151680 & -0.051 & 0.031 & 18 & 1.25 & 0.222 &   8.0 & 0.48 & 8.68 & 7.50 & 10.10 \\
MN48 & 5940216130049700480 & 0.323 & 0.135 & 12 & 1.02 & 0.469 &  77.9 & 1.38 & 2.82 & 2.00 & 4.40 \\
\hline
(HD80077) & 5325673208399774720 & 0.392 & 0.031 & 15 & 0.94 & 0.000 &   0.0 & 1.20 & 2.37 & 2.21 & 2.57 \\
(SBW1) & 5254478417451126016 & 0.128 & 0.030 & 17 & 0.87 & 0.000 &   0.0 & 1.58 & 6.01 & 5.10 & 7.26 \\
(Hen 3-519) & 5338229115839425664 & 0.042 & 0.030 & 17 & 1.03 & 0.000 &   0.0 & 1.61 & 9.57 & 7.72 & 12.17 \\
(Sher 25) & 5337418397799185536 & 0.072 & 0.033 & 19 & 0.91 & 0.000 &   0.0 & 1.59 & 7.88 & 6.39 & 10.03 \\
($\zeta^1$ Sco) & 5964986649547042944 & 0.713 & 0.242 & 12 & 0.94 & 0.947 & 297.5 & 1.36 & 1.51 & 1.01 & 2.74 \\
(HD 326823) & 5965495757804852992 & 0.743 & 0.053 & 11 & 1.05 & 0.000 &   0.0 & 1.41 & 1.30 & 1.22 & 1.40 \\
(WRAY 17-96) & 4056941758956836224 & 0.827 & 0.181 & 10 & 0.85 & 0.467 &  54.4 & 2.34 & 1.26 & 0.97 & 1.78 \\
(HD 316285) & 4057682692354437888 & 0.169 & 0.092 & 10 & 0.94 & 0.000 &   0.0 & 2.28 & 4.90 & 3.35 & 7.90 \\
(HD 168625) & 4097796621733266432 & 0.621 & 0.064 &  9 & 1.02 & 0.000 &   0.0 & 1.52 & 1.55 & 1.41 & 1.73 \\
(AS 314) & 4103870014799982464 & 0.624 & 0.052 & 11 & 0.96 & 0.000 &   0.0 & 1.75 & 1.54 & 1.42 & 1.68 \\
(MWC 314) & 4319930096909297664 & 0.191 & 0.042 & 13 & 0.91 & 0.000 &   0.0 & 1.34 & 4.36 & 3.67 & 5.33 \\
(W51 LS1) & 4319942771347742976 & 0.508 & 0.116 & 14 & 1.08 & 0.741 & 133.4 & 1.32 & 1.91 & 1.53 & 2.54 \\
(G79.29+0.46) & 2067716793824240256 & 0.180 & 0.139 & 15 & 1.06 & 0.648 & 150.3 & 0.79 & 3.09 & 2.29 & 4.36 \\
(CYG OB2 12) & 2067782734461462912 & 1.175 & 0.128 & 16 & 1.52 & 0.588 & 122.4 & 0.80 & 0.84 & 0.75 & 0.95 \\
\hline
\end{tabular}
%%%%%%%%%%%%%%%%%%%%%%%%%%%%%%%%%%%%%%%%%%%%%%%%%%%%%%%%%%%%%%%%%%%%%%%%%
%%%%%%%%%%%%%%%%%%%%%%%%%%%%
%\medskip
%$^a$ This distance is very uncertain because it may be a chance alignment of a foreground star, while the LBV may be highly reddened.
%$S$ is the proj h. \\
%%%%%%%%%%%%%%%%%%%%%%%%%%%%%%%%%%%%%%%%%%%%%%%%%%%%%%%%%%%%%%%%%%%%%%%%
\label{tab:tab2}\end{minipage}\end{table*}
\end{center}

%%%%%%%%%%%%%%%%%%%%%%%%%%%%%%%%%%%%%%%%%%%%%%%%%%%%%%%%%%%%%%%%%%%%%%%
\begin{center}
\begin{table*}\begin{minipage}{5.0in}%
    \caption{LBVs and LBV candidate (in parentheses) Gaia DR2 distances.}\scriptsize
\begin{tabular}{lcccccccccr} 
\hline
Name       &RA(deg)    &DEC(deg) &$d_{\text{Bayes}}$ (kpc) &
$d_{\text{low}}$ (kpc) & $d_{\text{high}}$ (kpc) & $d_{\rm lit}$ (kpc)
\footnote{$d_{\text{Bayes}}$ is the most likely distance to the LBV or LBV
candidate.  $d_{\text{low}}$ and $d_{\text{high}}$ give the highest 68\% density interval (HDI).  Details of the {\it Gaia} DR2 observations are in Table~\ref{tab:tab2}. $d_{\rm lit}$ is the nominal distance typically adopted in the literature (see text section 3).} \\
\hline
HR~Car & 155.72429 & -59.62454 & 4.37 & 3.50 & 5.72 & 5.20 \\
AG~Car & 164.04820 & -60.45355 & 4.65 & 3.73 & 6.08 & 6.00 \\
Wra~751 & 167.16688 & -60.71436 & 3.81 & 2.49 & 6.26 & 6.00 \\
Wd1 W243 & 251.78126 & -45.87477 & 1.78 & 0.83 & 4.16 & 4.50 \\
HD 160529 & 265.49594 & -33.50381 & 2.10 & 1.80 & 2.51 & 2.50 \\
HD 168607 & 275.31203 & -16.37550 & 1.46 & 1.31 & 1.65 & 2.20 \\
P Cyg & 304.44665 &  38.03290 & 2.17 & 0.98 & 4.25 & 1.70 \\
MWC 930 & 276.60514 &  -7.22165 & 4.46 & 2.89 & 6.93 & 3.50 \\
G24.73+0.69 & 278.48031 &  -6.97742 & 3.04 & 1.52 & 5.47 & 5.20 \\
WS~1 &  73.23917 & -66.68708 & 2.47 & 1.83 & 3.38 & 11.00 \\
MN48 & 252.40708 & -45.59980 & 2.74 & 1.44 & 5.18 & 5.00 \\
%HR~Car & 155.72429 & -59.62454 & 4.37 & 3.50 & 5.72 & 5.20 \\
%AG~Car & 164.04820 & -60.45355 & 4.65 & 3.73 & 6.08 & 6.00 \\
%Wra~751 & 167.16688 & -60.71436 & 4.32 & 3.38 & 5.82 & 6.00 \\
%Wd1 W243 & 251.78126 & -45.87477 & 1.01 & 0.84 & 1.26 & 4.50 \\
%HD 160529 & 265.49594 & -33.50381 & 2.10 & 1.80 & 2.51 & 2.50 \\
%HD 168607 & 275.31203 & -16.37550 & 1.46 & 1.31 & 1.65 & 2.20 \\
%P Cyg & 304.44665 &  38.03290 & 1.34 & 1.03 & 1.91 & 1.70 \\
%MWC 930 & 276.60514 &  -7.22165 & 7.01 & 5.10 & 9.73 & 3.50 \\
%G24.73+0.69 & 278.48031 &  -6.97742 & 5.21 & 3.42 & 7.94 & 5.20 \\
%WS~1 &  73.23917 & -66.68708 & 5.83 & 4.90 & 7.00 & 11.00 \\
%MN48 & 252.40708 & -45.59980 & 2.70 & 1.89 & 4.29 & 5.00 \\
%HR~Car & 155.72429 & -59.62454 & 4.41 & 3.66 & 5.51 & 5.20 \\
%AG~Car & 164.04820 & -60.45355 & 4.73 & 3.92 & 5.90 & 6.00 \\
%Wra~751 & 167.16688 & -60.71436 & 3.82 & 2.51 & 6.26 & 6.00 \\
%Wd1 W243 & 251.78126 & -45.87477 & 1.77 & 0.83 & 4.15 & 5.00 \\
%HD 160529 & 265.49594 & -33.50381 & 2.09 & 1.82 & 2.45 & 2.50 \\
%HD 168607 & 275.31203 & -16.37550 & 1.46 & 1.32 & 1.63 & 2.20 \\
%P Cyg & 304.44665 &  38.03290 & 2.17 & 0.98 & 4.25 & 1.70 \\
%MWC 930 & 276.60514 &  -7.22165 & 4.48 & 2.94 & 6.85 & 3.50 \\
%G24.73+0.69 & 278.48031 &  -6.97742 & 3.04 & 1.47 & 5.62 & 5.20 \\
%WS~1 &  73.23917 & -66.68708 & 2.48 & 1.94 & 3.21 & 11.00 \\
%MN48 & 252.40708 & -45.59980 & 2.74 & 1.44 & 5.18 & 5.00 \\
\hline
(HD80077) & 138.97824 & -49.97347 & 2.26 & 2.01 & 2.59 & 3.00 \\
(SBW1) & 160.08071 & -59.81940 & 5.11 & 4.03 & 6.78 & 7.00 \\
(Hen 3-519) & 163.49819 & -60.44564 & 7.12 & 5.45 & 9.65 & 8.00 \\
(Sher 25) & 168.78180 & -61.25488 & 6.28 & 4.82 & 8.52 & 6.30 \\
($\zeta^1$ Sco) & 253.49886 & -42.36204 & 2.52 & 1.12 & 4.97 & 2.00 \\
(HD 326823) & 256.72461 & -42.61104 & 1.27 & 1.17 & 1.40 & 2.00 \\
(WRAY 17-96) & 265.39765 & -30.11078 & 3.02 & 1.16 & 7.16 & 4.50 \\
(HD 316285) & 267.05848 & -28.01478 & 4.56 & 3.05 & 7.63 & 1.90 \\
(HD 168625) & 275.33145 & -16.37392 & 1.51 & 1.34 & 1.73 & 2.20 \\
(AS 314) & 279.85874 & -13.84646 & 1.50 & 1.36 & 1.68 & 8.00 \\
(MWC 314) & 290.39156 &  14.88245 & 3.93 & 3.15 & 5.11 & 3.00 \\
(W51 LS1) & 290.94849 &  14.61083 & 2.50 & 1.19 & 4.87 & 6.00 \\
(G79.29+0.46) & 307.92617 &  40.36639 & 1.87 & 1.07 & 3.27 & 2.00 \\
(CYG OB2 12) & 308.17065 &  41.24145 & 1.04 & 0.60 & 2.21 & 1.70 \\
\hline
\end{tabular}
%%%%%%%%%%%%%%%%%%%%%%%%%%%%%%%%%%%%%%%%%%%%%%%%%%%%%%%%%%%%%%%%%%%%%%%%%
%%%%%%%%%%%%%%%%%%%%%%%%%%%%
%\medskip\medskip
\\
%%%%%%%%%%%%%%%%%%%%%%%%%%%%%%%%%%%%%%%%%%%%%%%%%%%%%%%%%%%%%%%%%%%%%%%%
\label{tab:tab1}\end{minipage}\end{table*}
\end{center}

\begin{figure}
\includegraphics[width=\columnwidth]{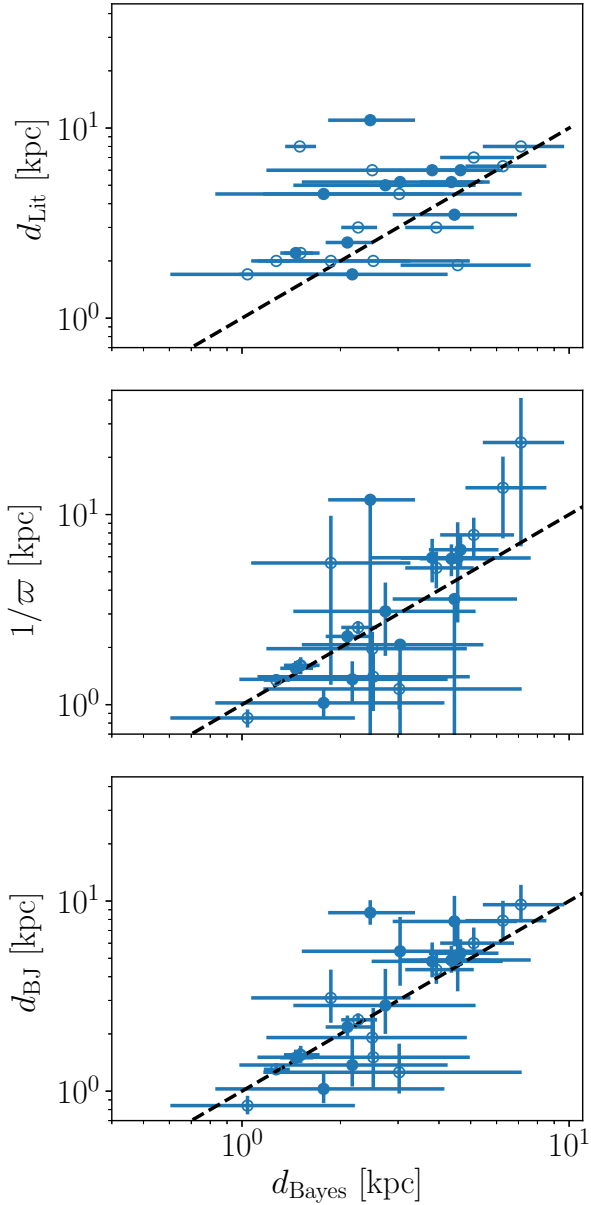}
\caption{Distances by Bayesian inference, $d_{\text{Bayes}}$ (Table~\ref{tab:tab1}) compared to: ({\it top})  literature distances, $d_{\text{Lit}}$, (Table~\ref{tab:tab1}), ({\it middle}) distances given by  $1/\varpi$ (Table~\ref{tab:tab2}), and ({\it bottom}) the Bailer-Jones Bayesian distances, $d_{BJ}$ \citep{bailer-jones2018} (Table~\ref{tab:tab2}). Filled circles represent the LBVs and open circles represent the LBV candidates.  For roughly half of the sample, $d_{\text{Bayes}}$ is significantly closer than the literature distances.  This has consequences for the inferred luminosities and masses for many LBVs. 
%%% Caption getting long...after this seems like analysis..don;t we say this in text?
%
%In general, $d_{\text{Bayes}}$ and $1/\varpi$ are consistent; they differ mostly due to the zero-point offset, which is included in $d_{\text{Bayes}}$.  Nine of the LBVs and LBV candidates have distances that are closer than the literature distances.  Given the 68\% HDI, one would expect 4.0 of the {\it Gaia} DR2 distances to be closer.  The Poisson probability that 9 are closer when 4.0 is expected is 1.3\%.  These closer distances affect the inferred luminosities and masses for a significant fraction of LBVs and candidates. The method for calculating $d_{\text{Bayes}}$ is similar to the method outlined by \citet{bailer-jones2018}.  The main differences are that $d_{\text{Bayes}}$ uses $\varpi_{\text{zp}}  = -0.05$ mas instead of $-0.029$ mas, and the uncertainty for $d_{\text{Bayes}}$ includes the astrometric excess noise and an uncertainty for the offset of $0.03$ mas.}   
%Generally, the two distances agree; for the largest distances, $d_{\text{Bayes}}$ tends to be closer, and for some objects the uncertainties are larger.  For some of the cases, the distribution of possible $d_{\text{Bayes}}$ are dominated by the prior rather than the observations.  In general, these cases that are dominated by the prior have uncertainties in $d_{\text{Bayes}}$ that are much larger than $d_{\text{BJ}}$.
}
\label{fig:AllThreeVsdBayes}
\end{figure}

\begin{figure*}
\includegraphics[width=5.2in]{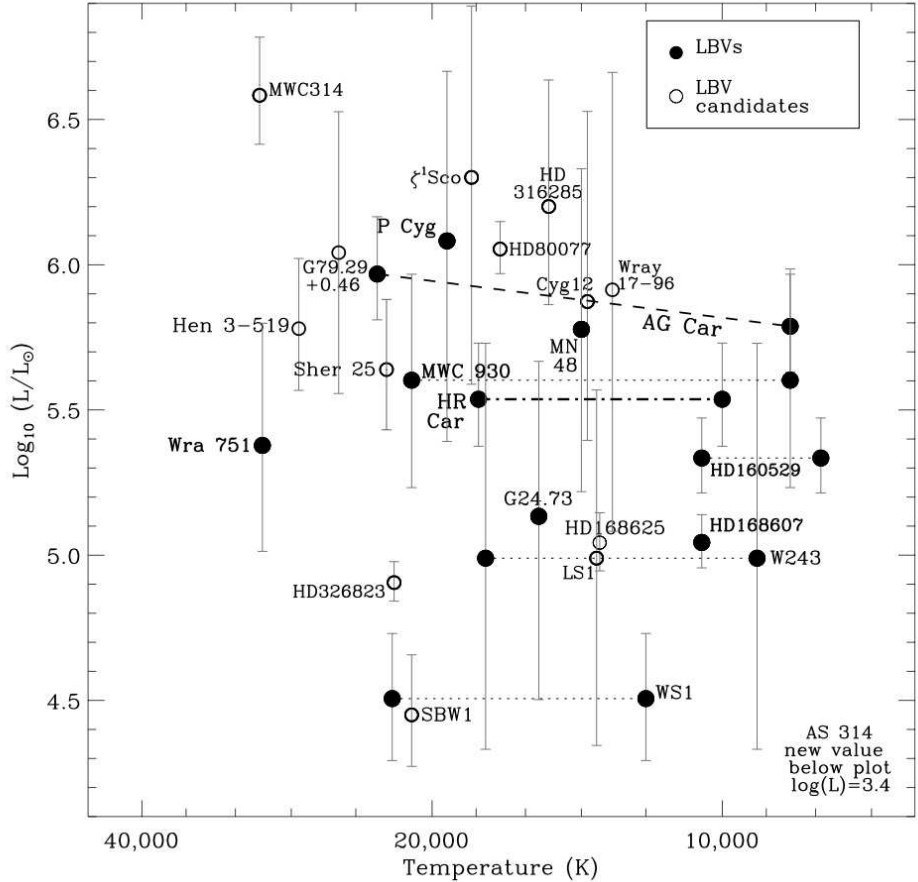}
\caption{The HR diagram showing only Galactic LBVs (filled circles) and 
  Galactic LBV candidates (unfilled circles) with their luminosities 
  scaled by the revised {\it Gaia} DR2 distances ($d_{\rm Bayes}$).  For this plot, we use the 
  new DR2 distances from Table~\ref{tab:tab1} (see text).  Here we do not 
  show the presumed location of the S~Dor instability strip, stellar 
  evolution model tracks, or any extragalactic LBVs.}
\label{fig:simple}
\end{figure*}

\begin{figure*}
\includegraphics[width=5.2in]{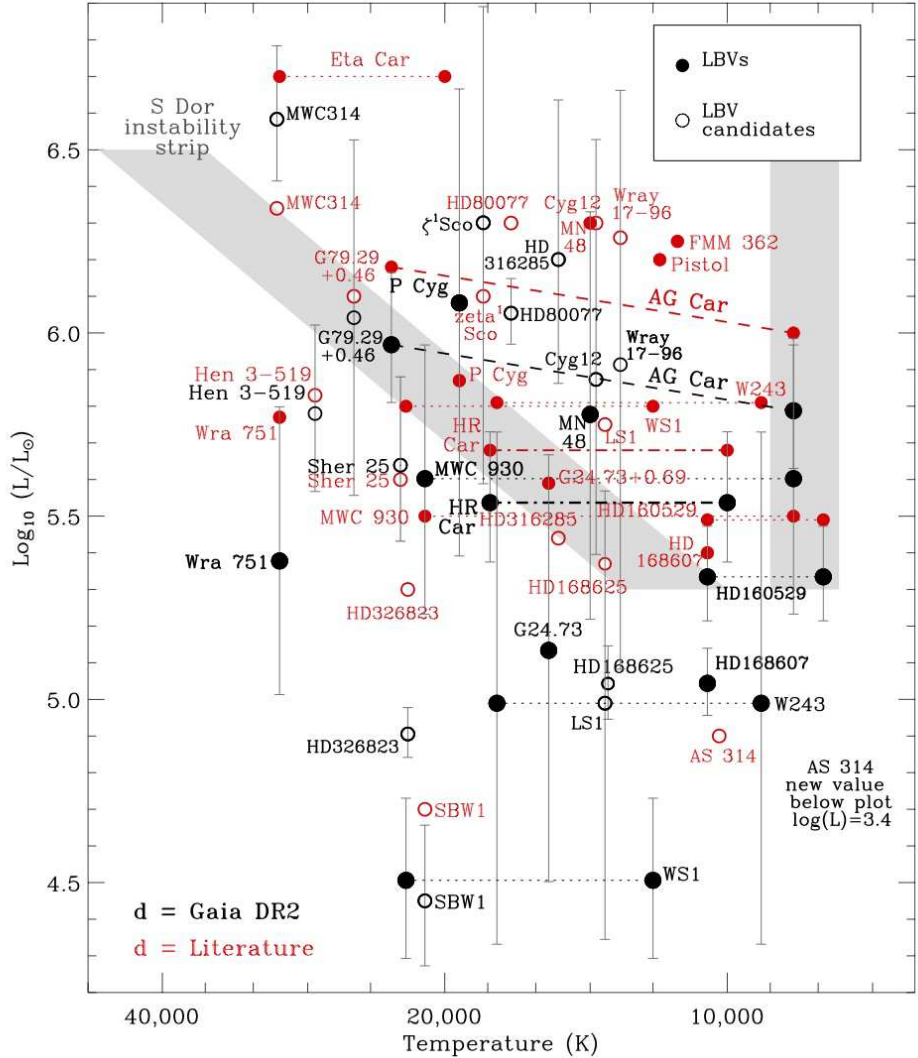}
\caption{Same as Figure~\ref{fig:simple}, but including positions of LBVs based on both the old literature distances (red) and those inferred from Gaia DR2 distances (black).  The S Dor instability strip and constant temperature eruptive LBV strip are also indicated, as in \citet{smith04}.}
\label{fig:oldnew}
\end{figure*}

\begin{figure*}
\includegraphics[width=6.5in]{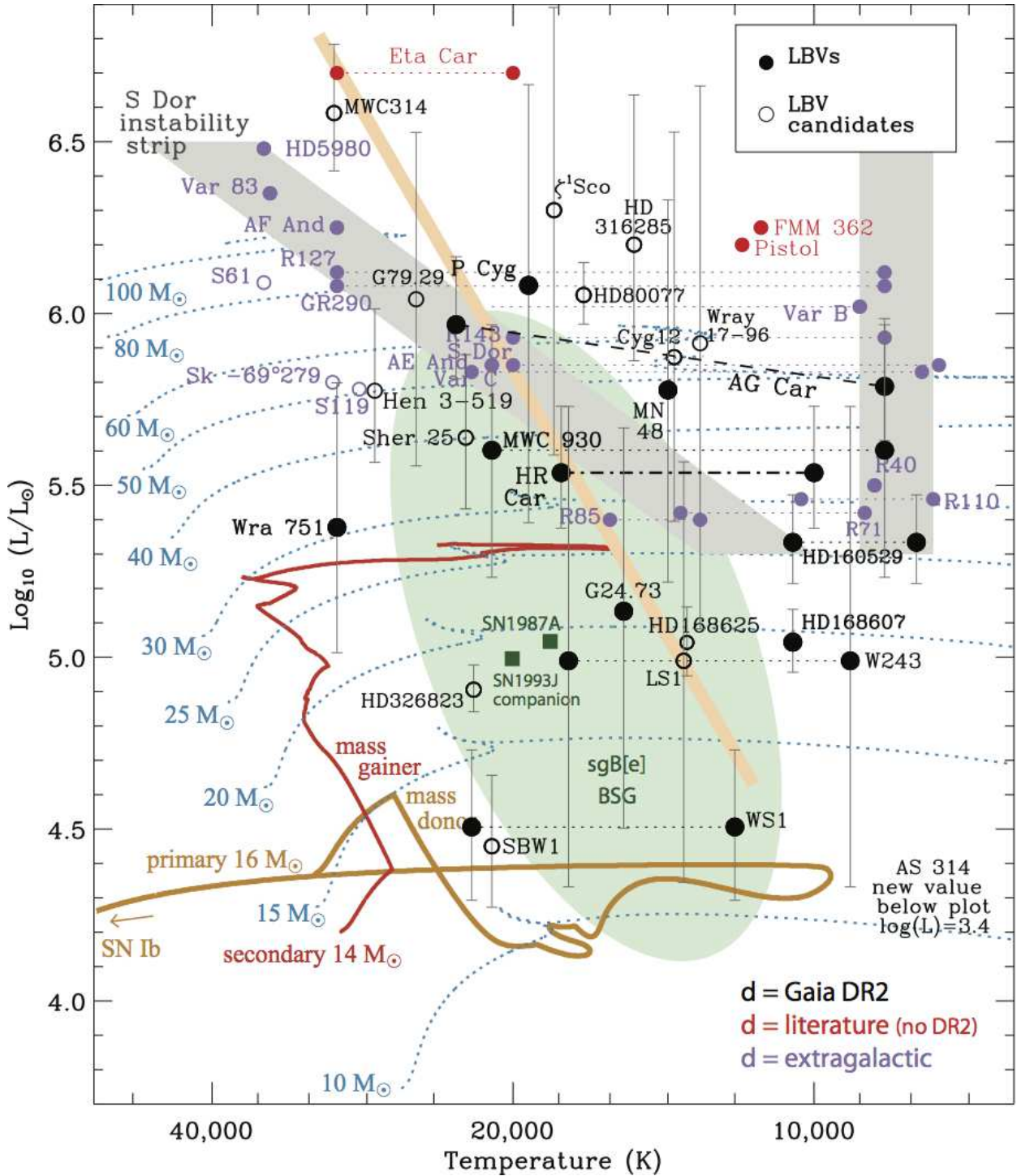}
\caption{The HR diagram with LBVs (filled circles) and LBV candidates
  (unfilled circles), adapted from a similar figure in \citet{st15}
  and \citet{ss17}.  Here, Galactic LBVs and LBV candidates are shown
  in black, with luminosities adjusted from old values as appropriate
  for the new {\it Gaia} DR2 distance.  For this plot, we use the DR2 distances 
  from Table~\ref{tab:tab1} (see text).  LBVs in nearby galaxies (LMC, SMC, M31, M33) are shown
  in light purple for comparison.  Locations of blue supergiants and
  B[e] supergiants, the progenitors of SN~1987A and SN~1993J, and some
  example stellar evolutionary tracks are also shown for comparison.
  The gray boxes show the locations of the temperature dependent S~Doradus 
  instability strip \citep{wolf89} and the constant temperature strip of 
  LBVs in outburst, as in \citet{smith04}.  The thinner orange line shows 
  the somewhat steeper S Doradus instability strip suggested by 
  \citet{groh09b} based on AG Car and HR Car (although it has been shifted 
  slightly here to accommodate their revised distances and luminosities from 
  DR2, and we have extrapolated over a larger luminosity range with the same 
  slope).  The single-star model tracks (blue) are from \citet{brott11} and the
  pair of binary system model tracks (red and pumpkin colored) is from \citet{lk14}. }
\label{fig:hrd}
\end{figure*}

\section{Distances for LBVs in Gaia DR2}

\subsection{Sample Selection}

We searched the {\it Gaia} DR2 \citep{gaia16,gaia18} database \citep{gaiaDR2}\footnote{\tt http://gea.esac.esa.int/archive/} for all known Galactic LBVs and
LBV candidates. As a convenient reference, we take the source list of 
Galactic LBVs and LBV candidates from the compilations by \citet{clark05} 
and \citet{st15}. To the list of \citet{clark05} we added SBW1 (candidate 
LBV), MWC~930, MN48, and WS1 (confirmed LBVs).  SBW1 was not listed in the compilation
of LBVs and candidates by \citealt{clark05} because it was discovered
later in 2007 \citep{sbw07}.  It has a ring nebula similar to that of
SN~1987A, and should be considered an LBV candidate for the same
reason that Sher~25 and HD~168625 have been included in past lists of cLBVs. 
MWC~930 was not included in the list by \citet{clark05}
because its LBV-like variability was discovered afterward in 2014
\citep{miro14}.  Our total sample of Galactic LBVs and cLBVs consists 
of 25 objects.

Of the LBV sources we checked, 4 confirmed LBVs ($\eta$ Car, GCIRS34W, FMM~362, 
and AFGL 2298) and 10 cLBVs (GCIRS 16NW, 16C, 16SW, 33SE, 16NE, the 
Pistol Star, WR102ka, LBV 1806-20, G25.520+0.216, and G26.47+0.02) did not have 
parallax values in Gaia DR2.  LBVs or LBV candidate stars in the vicinity of the 
Galactic Center are not listed in {\it Gaia} DR2 because they are visually obscured, 
including the Pistol Star, GCIRS 16NW, 16C, 16SW, 33SE, 16NE, 34W, etc.  We note 
that FMM~362 has an {\it almost} coincident {\it Gaia} source where the DR2 
parallax indicates a distance of only 1.6 kpc.  This consumed our attention for 
some time, but detailed examination of images shows that this 18th magnitude {\it Gaia} 
source is offset from FMM~362 by about 2$\farcs$25 and is likely to be a foreground 
K or M-type star. FMM~362 itself is highly obscured. Some objects with IR-detected 
shells are also not found for similar reasons, including IRAS~18576+0341 (AFGL~2248),
G25.520+0.216, and G26.47+0.02.  MN44 \citep{gvaramadze15} is a recent addition to the 
class of LBVs.  However, in {\it Gaia} DR2 it has a negative parallax and an extreme 
astrometric noise parameter.  Since there is no previous distance measurements (only 
example luminosities were given for assumed distances from 2 to 20 kpc), {\it Gaia} DR2 
does not improve the situation and we do not discuss MN44 further here; we await DR3 
for useful information on this source.

Also not included in {\it Gaia} DR2 is the very massive star $\eta$ Carinae.
Its parallax is not available, but in this case the absence is presumably due to complications
associated with its circumstellar Homunculus nebula.  Fortunately,
$\eta$ Car already has a reliable distance of 2.3 kpc based on the
expansion parallax of this nebula \citep{smith06}.

\subsection{Previous Literature Distances for LBVs}~\label{sec:lit}

Galactic distances are notoriously difficult to measure. This is especially true for LBVs because (a) their luminosity cannot be uniquely determined from stellar features, (b) they often undergo significant mass loss resulting in non-negligible circumstellar absorption, and (c) they sometimes exhibit peculiar velocities. As such, the previous literature estimates of the distances to the LBVs and cLBVs considered in this study arise from incredibly varied methods. 

%For simplicity, we adopt
For simplicity, we adopt the compilation of LBV+cLBV distances presented in the sample/review papers of \citet{naze12} and \citet{vg01} as our ``baseline'' literature distances for each star. Due to their presentation in a unified location, this distances have been widely used when assessing the luminosity of the population of Galactic LBVs. It is these distances, supplemented by measurements from \cite{Gvaramadze2012,Kniazev2016,smith07} for WS1, MN48, and SBW1, respectively, that are presented as $d_{\rm lit}$ in the second column of Table~\ref{tab:tab4} and the final column of Table~\ref{tab:tab1}.  However, in attempting to present the ``best'' distance value for each star, these compilations have sometimes averaged together measurements from multiple literature studies, and in all cases the full allowable distance range, formal errors (if given), and measurement method from original works have been obscured. Therefore in Table~\ref{tab:tab4} we additionally compile this information for publications whose measurements were utilized to compile our baseline distances. While not an exhaustive list of every published distance estimate, these values illustrate the typical range of distance measurements previously available for most galactic LBVs.

Table~\ref{tab:tab4} serves as a reminder of just how unreliable and heterogeneous previous distance and luminosity estimates have been for Galactic LBVs.  Once a distance was estimated in the literature, subsequent authors often engaged in more detailed study of the spectrum or variability of an LBV.  In doing so, it was common to adopt a representative or average distance for the sake of discussion.  These adopted values often propagated to subsequent papers as a "standard" value, but often without emphasizing or retaining the uncertainty in the original distance measurement (indeed, such bookkeeping becomes cumbersome for a caveat-loaded discussion, and many HR diagrams of LBVs in the literature have no error bars on the luminosity).  Nevertheless, the considerable uncertainty in distances has a profound impact on shaping our views of LBVs and their evolutionary scenarios.  Even when uncertainties in luminosity or distance were given, the quoted error bars might not include the true uncertainty.  For example, a distance estimate based on radial velocity and Galactic kinematics, or based on interstellar extinction, might yield a value with an error bar --- but without acknowledging that the method may not be valid for LBVs.  As noted above, Galactic rotation might not work as a distance estimate if LBVs have peculiar velocities because the have received a kick, for example, and the interstellar extinction method might not work if LBVs have their own circumstellar extinction, or if their local region has interstellar dust grains with anomalous reddening properties because they have been processed by strong UV radiation.  This serves to underscore the value of the new {\it Gaia} DR2 distances.  Even though some of the objects have quite large DR2 error bars, they are most often comparable to or smaller than previous estimates.  Moreover, they represent a single consistent method for all objects, and a direct (i.e., geometric) method that does not rely on sometimes dubious assumptions.

The most commonly utilized methods in literature distance estimates were cluster association plus main sequence fitting (7 objects), kinematics plus inferences from the Galactic rotation curve (7 objects), and extinction measurements coupled with distance-reddening relations (7 objects). Other methods occasionally invoked include comparison to other LBVs/assumption of a luminosity on the S Dor instability strip \citep[e.g.][]{Gvaramadze2012,miro00}, spectroscopic modeling \citep{najarro97}, maser proper motions \citep{Iamai2002,Schneps1981,Xu2009}, and spectroscopic parallax.

A majority of LBVs and LBV candidates considered have distances quoted in the literature whose full ranges span $>$3 kpc. In particular, 12 of the 25 stars have only either approximate distances reported (no associated errors) or upper/lower limits to their distances quoted. The most precise literature measurements available come from cluster association and main sequence fitting, with typical quoted errors of $\lesssim$0.3 kpc. Even these, however, suffer from the sometimes questionable assumption of membership.

Four stars in our sample have previous literature distance measurements that warrant specific discussion or mention:
\begin{itemize}
\item \emph{Wd1 W243:} The distance to Wd1 W243 has been derived based on its presumed association with the cluster Westerlund 1. While the baseline distance to Wd1 given in \cite{naze12,Ritchie2009} is 5.0 kpc, a range of distance estimates to Wd1 exist. \cite{clark05} initially quote a distance of $>$2 kpc and $<$5.5 kpc based largely on the luminosity of yellow hypergiants and the lack of identified Wolf Rayet (WR) stars in their data. \cite{Crowther2006} subsequently quote a distance of 5$^{+0.5}_{-1.0}$ kpc based on modeling of WR stars, although they note an observed dispersion of $\sigma\approx0.7$ in the distance modulus for Wd1, and caution that WR stars do not represent ideal distance calibrators. Most recently, \cite{Brandner2008} fit main-sequence and pre-main sequence evolutionary tracks to near-infrared data, finding both a more precise, and significantly lower, distance of 3.55$\pm$0.17 kpc. (In a separate paper, we discuss the revised cluster distance to Wd1 based on {\it Gaia} DR2; \citealt{mojgan18}.)
\item \emph{W51 LS1:} Similarly, the distance to cLBV W51 LS1 has been tied to the---surprisingly fraught---distance to the W51 complex (see \citealt{Clark2009a} and \citealt{Figueredo2008} for recent summaries). Early maser proper motion measurements gave relatively large distances of 6.1$\pm$1.3 kpc and 8.5$\pm$2.5 kpc \citep{Iamai2002,Schneps1981}. These were broadly consistent with kinematic measurement of $\sim$5.5 kpc \citep{Russeil2003}, leading \cite{Clark2009a} and \cite{naze12} to adopt a baseline distance to the cLBV W51 LS1 of 6 kpc. However, using spectroscopic parallax measurements of 4 O-type stars, \cite{Figueredo2008} find a much smaller distance of 2.0$\pm$0.3 kpc, a distance they acknowledge is difficult to reconcile with previous measurements. Most recently, updated trigonometric parallax measurements coupled with maser proper motions have given distances slightly lower than our baseline value, but still inconsistent with the spectroscopic parallax values:  5.1$^{+2.9}_{-1.4}$ kpc \citep{Xu2009} and 5.41$^{+0.31}_{-0.28}$\citep{Sato2010}.
\item \emph{HD168607/HD168625:} In the literature studies presented here, LBV HD168607 and cLBV HD168625 are both assumed to be members of the same stellar association, Ser OB1, and hence have the same distance. 
\end{itemize}

\subsection{New Gaia DR2 Data}

Table~\ref{tab:tab2} presents the {\it Gaia} DR2 data for the 25 LBVs and candidates (names for LBV candidates 
are given in parentheses) analyzed in this paper, and 
Table~\ref{tab:tab1} lists the distances that we infer from {\it Gaia} DR2
parallaxes. The last column of Table~\ref{tab:tab1} gives a ``baseline'' previously adopted distance from the literature (see Section~\ref{sec:lit} for further elaboration on this baseline value as well as a discussion of the methods, spread, and uncertainty in previous literature estimates). 

Table~\ref{tab:tab2} includes the {\it Gaia} DR2 data and distance
  estimates from \citet{bailer-jones2018}.  The first two columns
  present the parallax, $\varpi$, and the theoretical uncertainty in
  the parallax, $\sigma_{\varpi}$.  One
  could simply report $d = 1/\varpi$, and one could use
  $\sigma_{\varpi}$ to calculate an uncertainty for the distance.
  However, such a calculation would not fully account for all of the calibration issues or extra
  sources of uncertainty.  Furthermore, the uncertainty for many objects
in DR2 is quite large.  In fact, the uncertainty can be so large that
the parallax can be negative.  In these circumstances, inferring the
distance and uncertainty from $d = 1/\varpi$
is either inaccurate or impossible.  A preferred solution for all of these
issues is to infer the distance using Bayesian inference \citep{luri18}. 

The fifth and sixth columns present the
  excess astrometric noise, $\epsilon$, and the excess astrometric
  noise significance, $D$.   See \citet{lindegren2012} for a thorough discussion of the
  astrometric noise.  In short, the excess astrometric noise
  represents variation in the astrometric data beyond the standard
  astrometric solution.  The standard astrometric solution is composed
of five parameters: two sky coordinates, two proper motion
coordinates, and a parallax.  Any unknown calibration issues or extra
motions due to binaries, etc., lead to excess astrometric noise.  $D$
is a measure of the significance of the excess astrometric noise. 
%In an ideal situation with no true excess noise, there is no excess noise in only about half of the measurements.  
Even in the ideal situation with no true excess noise,
random noise produces excess noise in about half the measurements.  In
the ideal situation, 98.5\% of measurements have a significance of $D
< 2$.  Column seven of Table~\ref{tab:tab2} shows significant excess noise for several of the
LBV and candidate sources.  Therefore to properly infer the
distribution for the distance, one must include the excess noise.

Columns 10$-$12 of Table~\ref{tab:tab2} give the Bayesian inference for the distances by
  \citet{bailer-jones2018}.  Unfortunately, these distance estimates do not
  include the excess noise, (WELL, THEY DO - JUST NOT IN A GOOD WAY!) 
  so we merely report them for comparison.
  To obtain the Bayesian-inferred distances, we searched the catalogue
  at {\tt http://gaia.ari.uni-heidelberg.de/tap.html} which reports
  geometric distances inferred from {\it Gaia} DR2 parallaxes
  \citep{bailer-jones2018}.  Column nine gives an estimate for the
  length scale of field stars in the direction of the LBV or
  candidate.  This length scale is a parameter in the prior of the
  \citet{bailer-jones2018} estimate.  $d_{\text{BJ}}$ is the mode of
  their posterior, and $d_{\text{BJ,low}}$ and $d_{BJ,high}$ give the
  upper and lower bounds for the highest density 68\% confidence
  interval.  \citet{bailer-jones2018} used a zero-point offset of
  $\varpi_{\text{zp}} = - 0.029$ mas for their inference.  However,
  subsequent work indicates that the zero-point has significant
  variation \citep{lindegren2018,riess2018,stassun2018,zinn2018}.

\subsection{New Bayesian-inferred distances}

The primary goal here is to use Bayesian inference to estimate the distance given by data in {\it Gaia} DR2.
To infer the distance to any star, the posterior distribution for the distance, $d$, and the zero point, $\varpi_{\text{zp}}$ is
\begin{multline}
\label{eq:posterior1}
P(d,\varpi_{\text{zp}}|\varpi,\sigma_{\varpi},\epsilon,\mu_{\text{zp}},\sigma_{\text{zp}},\ell)
\\ \propto \mathcal{L}(\varpi|d,\sigma_{\varpi},\epsilon,\varpi_{\text{zp}}) \times
P(\varpi_{\text{zp}} | \mu_{\text{zp}}, \sigma_{\text{zp}}) \times
P(d|\ell) \, .
\end{multline}
$\varpi$ and $\sigma_{\varpi}$ are the parallax and expected uncertainty, respectively, 
$\epsilon$ is the excess astrometric noise, and $\mu_{\text{zp}}$ and $\sigma_{\text{zp}}$ 
are the mean and variance of the zero point parallax.  $\mathcal{L}(\varpi|d,\sigma_{\varpi},\epsilon,\varpi_{\text{zp}})$ is
the likelihood of observing parallax $\varpi$ given the model parameters.  
$P(\varpi_{\text{zp}} | \mu_{\text{zp}},\sigma_{\text{zp}})$ is the
distribution of zero points.  $P(d|\ell)$ is the prior of observing
$d$ given the galactic length scale of stars in the FOV.

The parameter $\sigma_{\varpi}$ is the expected theoretical uncertainty in the parallax, 
but it does not represent the full uncertainty.  \citet{lindegren2018} 
found that the empirical uncertainty is 1.081 times larger 
than the expected value.  For many objects there is also an excess 
astrometric noise, $\epsilon$.  Formally, $\epsilon$ represents excess 
noise in the entire astrometric solution, not just the parallax. The excess noise absorbs the residual in the astrometric solution.  It is determined by adding $\epsilon$ in quadrature to the position uncertainty in the denominator of the $\chi^2$ minimization formula.  Minimizing $\chi^2$ finds a simultaneous solution for the five astrometric parameters and the excess noise.  The uncertainties of the astrometric solution are calculated using standard error propogation.  Therefore, the uncertainty in the DR2 parallax includes the excess noise.  However, the procedure for including the excess noise assumes that the excess noise is gaussian and uncorrelated among the observed positions.  If the excess noise is a correlated systematic residual, then the excess noise model in the {\it Gaia} pipeline would not capture the true behavior of the residual.  For example, when assuming that the excess noise is uncorrelated, the uncertainty in the parallax would go down by $\sqrt{N}$, where $N$ is the number of observations.  On the other hand, a systematic uncertainty may not reduce with added observations. 

Without 
knowing the source of the excess noise, it is difficult to determine whether the excess noise should be modeled as correlated or uncorrelated noise.  \citet{mojgan18} measured the empirical uncertainty distribution for the star cluster Westerlund 1 in the same way that \citet{lindegren2018} measured the empirical uncertainty distribution for all quasars in {\it Gaia} DR2.  Whereas the empirical uncertainty is 1.081 times larger for the quasars, \citet{mojgan18} found that most of the stars have a large astrometric excess noise and the empirical uncertainty for the stars in Westerlund 1 is 1.6 times larger than the DR2 solution.  This may suggest that the assumption of uncorrelated excess noise is inconsistent with the data.  Instead, the excess noise may not be random noise but a correlated residual.

Since this paper analyzes individual stars, it is impossible to empirically measure the uncertainty for each.  Yet, several LBVs have large astrometric excess noise, and it is likely that the excess noise contribution to the parallax uncertainty is underestimated.  W243 illustrates this point.  W243 is in Westerlund 1, which has a distance of $3.2 \pm 0.4$ kpc \citep{mojgan18} as determined by a large statistical sample of DR2 parallaxes of cluster members. Yet, the DR2 distance to the individual star W243 is $1.03^{+0.24}_{-0.17}$ kpc.  Using the DR2 uncertainty, $0.24$ kpc, this would be 9 sigma from the statistically inferred distance to Westerlund 1.  Even using the statistical uncertainty of $0.4$ kpc from \citet{mojgan18}, the {\it Gaia} DR2 estimate for W243 is still 5 sigma from the statistical cluster estimate.  Clearly, the parallax uncertainty in DR2 is underestimated in some cases.

Given the above problems with the astrometric excess noise, and our goal of estimating distances for individual objects, we choose to adopt a conservative uncertainty estimate. In our analysis, we found that making different assumptions nevertheless led to robust results for sources with small or zero $\epsilon$, but for sources with significant astrometric noise, the resulting distance could jump by much more than the DR2 uncertainty.   Hence, the excess residual that contributes to the parallax should be increased for such sources. In the absence of further information, we adopt the most conservative approach, 
and add the full excess noise, $\epsilon$ in quadrature with the theoretical 
parallax uncertainty, $\sigma_{\varpi}$:
\begin{equation}
\sigma^2 = (1.081\sigma_{\varpi})^2 + \epsilon^2 \, .
\end{equation}
While this yielded rather large uncertainty for some sources, we found that this uncertainty encompassed the full variation we had found in derived distances under different assumptions in the analysis.  With the larger uncertainty included this way, we achieved consistent results.  In many cases, these large uncertainties in the resulting distances for sources with large $\epsilon$ are still an improvement over estimates in the literature.

Inferred distances are also affected by the adopted zero point offset in the astrometric solution for the parallax.  $\mu_{\text{zp}}$ and $\sigma_{\text{zp}}$ are the mean and variance for the 
zero point parallax.  In the initial astrometric solution for DR2, \citet{lindegren2018} 
used quasars to quantify the zero point.  They found an average of $-$0.029 mas.  
In addition, they noted significant variation in the zero point as a function of sky
position and other possible parameters such as brightness and color.  They did not
quantify the variation in zero point; rather they encourage users of
{\it Gaia} DR2 to model their problem-specific zero point.
\citet{riess2018} use {\it Gaia} DR2 parallaxes to calibrate Cepheid
distances, and infer a zero point of $-$0.046 ($\pm$0.013) mas.
\citet{zinn2018} compare the distances inferred from astroseismology
to infer a zero point of $-$0.0528 ($\pm$0.0024) mas. \citet{stassun2018}
use distances derived from eclipsing binaries to infer a zero point of
$-$0.082 ($\pm$0.033) mas.  These are all estimates for the mean zero
point $\mu_{\text{zp}}$; \citet{lindegren2018} note a spatial
variation of the offset of about $\pm$0.03 mas in the direction of the
LMC.  Therefore, we assume that $\sigma_{\text{zp}} = 0.03$.  Unfortunately,
determining the zero point specifically for LBVs is difficult.  They typically lie
in the plane of the Galaxy, where there are no observable quasars, and
there are no accurate distance measures for LBVs.  Therefore, our best
estimate for the zero point is the mean of the above four
investigations, $\mu_{\text{zp}} = 0.05$
mas, and we use the spatial variation from \citet{lindegren2018} for
$\sigma_{\text{zp}} = 0.03$.  These measures of the zero-point are far
from ideal, but they are a good first attempt.

The likelihood and distribution for the zero points are as
follows:  The likelihood for observing parallax $\varpi$ is
\begin{equation}
\label{eq:likeli}
\mathcal{L}(\varpi|d,\sigma_{\varpi},\epsilon,\varpi_{\text{zp}}) =
\frac{1}{\sqrt{2 \pi} \sigma} \exp \left [ \frac{-(\varpi - 1/d -
  \varpi_{\text{zp}})^2}{2 \sigma^2} \right ] \, .
\end{equation}
\citet{lindegren2018} show that the zero point has significant
  variation as a function of sky location.  Therefore, we model the
  distribution of zero points as
\begin{equation}
\label{eq:zpdistribution}
P(\varpi_{\text{zp}}| \mu_{\text{zp}},\sigma_{\text{zp}}) =
\frac{1}{\sqrt{2 \pi} \sigma_{\text{zp}}} \exp \left [\frac{-(\varpi_{\text{zp}} -
  \mu_{\text{zp}})^2}{2\sigma_{\text{zp}}^2} \right ] \, . 
\end{equation}
$P(d|\ell)$ represents the prior for $d$ based upon the Galactic distribution of stars and dust extinction.  Consider an image populated with Galactic stars.  The total number of stars in the image is given by $N = {\rm FOV} \int n r^2 dr$, where FOV is the field of view in square radians, and $n$ is the number density of stars.  If $n$ is constant, then any random star in the image is drawn from a probability distribution of $P(r) \propto r^2$.  In the presence of dust extinction, this distribution will be attenuated by $\exp{(-r/\ell)}$, where $\ell$ is an effective optical depth for extinction.  For these reasons, \citet{bailer-jones2018} use the following prior when calculating the geometric distance to stars in {\it Gaia} DR2:
\begin{equation}
\label{eq:prior}
P(d|\ell) = \frac{1}{2\ell^3} d^2 \exp(-d/\ell) \, ,
\end{equation}
where $\ell$ is the attenuation length and depends upon the Galactic 
coordinates $(l,b)$.  This prior has a mode at $2\ell$.

To find the posterior distribution for only the distance, we marginalize 
eq.~(\ref{eq:posterior1}) over the nuisance parameter, the zero point, 
$\varpi_{\text{zp}}$.  In practice, this marginalization involves a 
convolution of the likelihood, eq.~(\ref{eq:likeli}), with the distribution 
for zero points, eq.~(\ref{eq:zpdistribution}). The convolution of two 
Gaussian distributions is analytic and so is the final posterior distribution.
\begin{multline}
\label{eq:posterior2}
P(d|\varpi,\sigma_{\varpi},\epsilon,\mu_{\text{zp}},\sigma_{\text{zp}},\ell) \\
\propto
\frac{1}{\sqrt{2 \pi (\sigma^2 + \sigma^2_{\text{zp}})}} \exp \left [\frac{-(\varpi - 1/d -
  \mu_{\text{zp}})^2}{2 (\sigma^2+\sigma^2_{\text{zp}})} \right ]
\\ \times
\frac{1}{2\ell^3} d^2 \exp(-d/\ell) \, .
\end{multline}

The geometric distances in columns 4-6 of Table~\ref{tab:tab1} are the mode 
(column 4) and the highest density 68\% interval (HDI, columns 5 \& 6) for 
this posterior distribution.  The seventh column in Table~\ref{tab:tab2} 
gives the attenuation scale, $\ell$, in the prior.  These were calculated 
using the same technique as in \citet{bailer-jones2018}.\footnote{The python code to 
calculate these distances is available at the following GitHub repository: {\tt 
https://github.com/curiousmiah/Gaia\_Distances}.}   The posterior 
distribution seamlessly handles both accurate and inaccurate parallax measurements.  
In the limit of an accurate parallax measurement, the width of the posterior 
will be dominated by the Gaussian and consequently $\sigma$.  In the limit of 
very inaccurate parallax measurements, the mode and uncertainty will be dominated 
by the prior.  Therefore, in the inaccurate cases, the most likely distance will 
be $2\ell$.

Since LBVs may be atypical in their distance distribution, it is not clear that the
\citet{bailer-jones2018} prior is the most appropriate prior to use.  LBVs tend
to be quite bright, so the distance scale, $\ell$, should probably be
larger, and it is not clear that LBVs should trace the general
population of stars.  The prior mostly affects the result of
those with very uncertain parallaxes.

%Since the uncertainty is quite
%large, the impact of the prior is fairly inconsequential.

Figure~\ref{fig:AllThreeVsdBayes}
%,~\ref{fig:BayesVsBailer},~and~\ref{fig:Bayeslit} 
compares all four
distance estimates: our Bayesian-inferred distance ($d_{\rm Bayes}$) is compared to 
({\it bottom panel}) the Bailer-Jones Bayesian estimate, then ({\it middle panel}) the simple $1/\varpi$ calculation ($d_{\varpi}$), and ({\it top panel}) literature
distances ($d_{\rm lit}$).  In all panels, the filled circles
represent LBVs and the open circles represent the LBV candidates.  

The bottom panel of Figure~\ref{fig:AllThreeVsdBayes} compares the mode and 68\% HDI of
eq.~(\ref{eq:posterior2}), $d_{\text{Bayes}}$ with the mode and HDI of
\citet{bailer-jones2018}, $d_{\text{BJ}}$.  The primary differences
are that the posterior distribution for $d_{\text{Bayes}}$ includes a more conservative prescription for 
the excess astrometric noise, a slightly more negative zero point, and a 
variation in the zero point.  The zero point in the $d_{\text{BJ}}$ 
distances is $-$0.029~mas, and the zero point for $d_{\text{Bayes}}$ is 
$-$0.05 ($\pm$0.03)~mas.  As a result, the uncertainties for 
$d_{\text{Bayes}}$ are larger when the excess noise is non-zero. Because 
the zero point for $d_{\text{Bayes}}$ is slightly more negative, those 
distances tend to be smaller than $d_{\text{BJ}}$.

The middle panel of Figure~\ref{fig:AllThreeVsdBayes} compares $d_{\text{Bayes}}$ with $d_{\varpi} = 1/\varpi$.
Naturally, both methods are roughly consistent; the derivation of $d_{\text{Bayes}}$ uses $\varpi$ after all.  However, there are some noteworthy differences.  At distances larger than $\sim$4 kpc, the
$d_{\varpi}$ distances are systematically larger than $d_{\rm Bayes}$.
This systematic discrepancy is mostly due to the zero-point offset.
For some, the prior distribution dominates the posterior for $d_{\text{Bayes}}$; for these distance estimates, $d_{\text{Bayes}}$ tends to have much larger mode and uncertainty.

The top panel of Figure~\ref{fig:AllThreeVsdBayes} compares $d_{\rm Bayes}$ with the baseline previous literature estimates, $d_{\rm lit}$, described in Section~\ref{sec:lit}.  On average, the literature distances are larger than the new inferred distances; the average literature distance is $4.1 \pm 0.6$ kpc, and the average $d_{\text{Bayes}}$ distance is $3.1 \pm 0.3$ kpc.  For about half of the objects,
the two estimates are consistent.  For most of the other half,
the literature distances significantly overestimate the distance.
To be more quantitative, there are 25 LBVs and candidates.  11 of
the literature distances are outside of the 68\% confidence intervals
(CIs). In particular, 2 are below the CIs, and 9 are above.  The
expectation is that roughly 16\% should be below and 16\% should be
above; 16\% of 23 is 4.  The Poisson probability of 2 below when
the expected value is 4 is 15\%, which is roughly consistent.
The Poisson probability of 9 above is 1.5\%.  As a population, there 
is a tension between the literature distances and the {\it Gaia} DR2 distances 
in that the literature distances tend to be too large.  This has a 
significant impact on the inferred luminosities and masses of LBVs.
%Possibly offer assessment of which methods are most discrepant? Or do that below. 

Figure~\ref{fig:simple} shows a revised HR diagram for LBVs using updated
distances from {\it Gaia} DR2 from Table~\ref{tab:tab1}.   To construct this, we adopted previously 
published values of $T_{\rm eff}$ and $L_{\rm Bol}$ compiled from the literature 
\citep{ss17,st15,clark05,naze12} with their associated previous distances (see Section 3), and we simply 
scaled the bolometric luminosities appropriate to the new DR2 distances\footnote{Recall that 
determinations of T$_{\rm{eff}}$ from spectroscopic analysis have negligible dependence on the
distance (Section~\ref{sec:intro}).}.  Figure~\ref{fig:oldnew} shows a similar HR diagram with the new DR2 
values compared to previous literature estimates.   Some LBVs changed little and others changed dramatically.  
LBV positions based  on previous literature values are plotted in red, and those with 
their $L_{\rm Bol}$ scaled by the new {\it Gaia} DR2 distances are plotted in
black.   Figure~\ref{fig:hrd}  then shows these same new values, but superposed with additional information 
for context, including  
extragalactic LBVs in nearby galaxies, the previously proposed S~Doradus instability 
strip, locations of B supergiants and B[e] supergiants, a few SN progenitors, and 
representative stellar evolution model tracks.    Extragalactic LBVs in the LMC, SMC, M31, and M33 are plotted
in light purple for comparison. Representative single-star and binary
evolution tracks are included for comparison, as in earlier versions
of this figure by \citet{ss17} and \citet{st15}.  These model tracks
are from \citet{brott11} and \citet{lk14}.

\subsection{The Significance of the Zero-point and the Lower Distances}

The {\it Gaia} DR2 parallaxes provide the largest collection to date
of LBV and LBV candidate distances that are measured in a uniform and direct way.  As an ensemble,
these distances are significantly lower than previous literature
distances.  In total, there are 25 LBVs and candidates, 11 have
literature distances that are larger than the 68\% HDI.  The {\it Gaia} DR2
distances are closer than previous estimates for two reasons.  First, on
average, the measured parallaxes are larger.  Second, there is a
significant negative zero point offset for {\it Gaia} DR2.  In other words, the true
parallax is larger by 0.05 mas.
%Distances with zero offset, but the uncertainty is still 0.02
%AG~Car & 164.04820 & -60.45355 & 5.99 & 4.94 & 7.51 & 6.00 \\
%HD 168607 & 275.31203 & -16.37550 & 1.57 & 1.42 & 1.76 & 2.20 \\
HD168607 is an example for which the {\it Gaia} DR2 distance is closer even if one ignores the zero-point offset.  The previously reported distance to HD~168607 is 2.2 kpc.  Including the zero-point offset, the {\it Gaia} DR2 distance is $1.46^{+0.15}_{-0.14}$ kpc.  Without applying the zero-point
offset, the {\it Gaia} DR2 distance is $1.57^{+0.19}_{-0.15}$ kpc.

AG~Car is an example for which the offset changes the consistency between the {\it Gaia}
DR2 distance and the previous estimate.  The
previous distance estimate was 6 kpc.  With a zero point offset the
Gaia DR2 distance is $4.73^{+1.17}_{-0.81}$ kpc, and without the zero
point, it is $6.0^{+1.5}_{-1.1}$ kpc.  With the offset, there is a mild
tension between the {\it Gaia} DR2 distance and the previous distance
esimate.  Without the offset, both distances are consistent.

Collectively, if one omits the offset, then the number of objects
with previous distances that are outside the 68\% HDI reduces from 11 to
10.  The Poisson probability of 8 when 4 is expected is 3\%, while
for 9 it is 1.3\%. Therefore, the tension only slightly depends upon the reliability of the zero-point 
estimate.  It is difficult or impossible to measure the offset for each 
LBV directly.  There are few accurate independent distance estimates for 
LBVs in the Galaxy, and because they reside in the Galactic plane, there 
are no bright background quasars to provide the zero-point calibration.  
Based upon the estimates of four independent groups, the global average offset 
is consistently measured to be about $-$0.05~mas.  The average offset for 556,869 quasars is $-$0.029 mas
\citep{lindegren2018}; the average offset for 50 Cepheid variables
is $-$0.046 ($\pm$0.013)~mas; the average offset of $\sim$3500 giants with
astroseismology is $-$0.0528 ($\pm$0.0024)~mas, and the average offset
for 89 eclipsing binaries is $-$0.082 ($\pm$0.033)~mas.  The mean of these is
$-0.05$ and the standard deviation is $0.02$ mas.  This represents the
global average, but \citet{lindegren2018} note a significant spatial
variation in the offset.  In the direction of the LMC the variation
is about $\pm$0.03~mas.

%\mrd{NB: Full table can be found \href{https://docs.google.com/spreadsheets/d/1VI7KjJ1Zcpn0F6lkfQgT2ZV7lo3n0feAQbPNAASYqdQ/edit#gid=0}{here}.} 

\section{NOTES ON INDIVIDUAL LBVs}

Here we provide notes on each of the 25 LBVs and candidates discussed in this work and the impact of the new {\it Gaia} DR2 distances.  In this sample of 25 LBVs and candidates, 11 have {\it Gaia} DR2 posterior distances for 
which the 68\% HDI is closer than the literature distances.  This represents almost half of 
the LBV and cLBV sample, even though the expectation is that only 4 should have 
closer distances.  The Poisson probability of having 9 closer when 4 are expected is 
1.3\%.  Some of these distance estimates changed by a large factor.

%We have divided the sample into three categories based on their {\it Gaia} DR2 and previous literature distances: stars with quite uncertain {\it Gaia} DR2 distances (9 objects), stars whose distances did not change much (5 objects), and stars where {\it Gaia} DR2 implied significant changes to their luminosity (11 objects). For simplicity, we distinguish between the latter to categories based on whether the ``baseline'' literature distance described in Section~\ref{sec:lit} falls inside or outside the 68\% confidence region of our new {\it Gaia} DR2 distances (Table~\ref{tab:tab1}). 

\subsection{Confirmed LBVs}

{\it HR Carinae:} 
The {\it Gaia} DR2 Bayesian-inferred distance (Table~\ref{tab:tab1}) 
is 15\% closer than its traditional value.  Its luminosity therefore moves down by 28\% 
on the HR Diagram.  The {\it Gaia} DR1 parallax yielded a smaller distance around 2 kpc as 
well, although with a large uncertainty, so \citet{ss17} did not advocate a revision
to the traditional distance. The new value is consistent within the uncertainty of DR1, 
although the distance from DR2 is more precise.  Here, the new distance is somewhat lower 
than $d_{\rm lit}$ and the star somewhat less luminous, but marginally consistent with the 
traditional value within the uncertainty. Therefore, the {\it Gaia} DR2 distance for HR Car 
does not significantly alter our interpretation of this objects parameters \citep{groh09b}.

{\it AG Carinae:}  As for HR Car, the {\it Gaia} DR2 distance derived in Table~\ref{tab:tab1} 
is slightly closer than the traditional value.  Adopting the most likely value in 
Table~\ref{tab:tab1} (plotted in Figure~\ref{fig:hrd}), AG~Car is about 20\% closer (moving 
from 6 kpc to 4.7 kpc) and is therefore about 40\% less luminous than previously determined 
from detailed analysis by \citet{groh06,groh09}.  However,the large uncertainty means that 
there is only a weak tension between the {\it Gaia} DR2 and literature distance. In fact, 
accounting for the errors in the original kinematics and reddening-distances measurements 
\citep{Humphreys1989,Hoekzema1992}, there is overlap with the upper end of the Gaia DR2 
68\% HDI. This is in contrast to the much closer distance inferred from 
the {\it Gaia} DR1 parallax, which was reported to be about 2 kpc \citep{ss17}.  
That closer distance was evidently an underestimate, although the range of distances 
indicated by the DR1 uncertainty of 1.3 to 3.7 kpc are marginally consistent with the DR2 
68\% HDI range found here of 3.7 tp 6.1 kpc. A much closer distance for AG~Car would have 
been surprising and problematic, since it would have moved AG Car -- a prototypical 
classical S~Dor variable -- to be far below the S~Dor instability strip that it helped to 
define, and far from its near twin R~127 in the LMC.  {\it Gaia} DR2 values suggest that 
more modest revision to its traditionally assumed distance is needed.  The new DR2 value 
is consistent with the traditional S Dor strip, although its implied effective initial 
mass (compared to single-star models) would be around 60-70 $M_{\odot}$ instead of 
$\sim$90~$M_{\odot}$.

{\it Wra 751:} The {\it Gaia} DR2 distance for this LBV in Table~\ref{tab:tab1} 
moves it from a baseline literature distance of $\sim$6 kpc to $3.8^{+2.4}_{-1.3}$ kpc. 
While a significant nominal correction, this is just barely consistent with the old value 
within the uncertainty, and has more overlap with the original distance measurements of 
\cite{vanGenderen1992} who simply quote limits of $\gtrsim$4$-$5 kpc. With the more reliable 
uncertainties of the {\it Gaia} DR2 distance, this provides a stronger case that Wra~751 
is significantly below or hotter than the S~Dor instability strip on the HR Diagram.  If 
Wra~751 really is off the S~Dor instability strip, then this is interesting because it is 
now considered a confirmed LBV, due to the development of apparent S~Dor variability 
\citep{sterken08,clark05}.   This is also of interest because Wra 751 has been associated 
with a presumed host cluster \citep{pasquali06}. It therefore bucks the trend 
that LBVs statistically avoid O-star clusters \citep{st15}.  Although Wra~751 is in a 
cluster, the cluster's age and turnoff mass do not agree with those of the LBV if it has 
evolved as a single star.  From its position on the HR Diagram, we would expect Wra~751
to have an effective single-star initial mass around 30-40 $M_{\odot}$.  As noted by 
\citet{pasquali06}, however, the earliest spectral type main sequence star in the host 
cluster is O8 V, translating to a cluster turnoff mass around 20 $M_{\odot}$.  This is 
consistent with results from LMC LBVs, where it has been noted that in the few
cases when LBVs are seen in a cluster, they seem too young and massive for their 
environment \citep{st15}.  This, in turn, reinforces ideas about the role of binary 
evolution and rejuvenation that make LBVs massive blue stragglers 
\citep{st15,smith16,mojgan17}. Wra~751 would be an excellent case study for investigating 
how its proper motion compares to those of the cluster members.

{\it Wd1 W243:} This is a special case because it has a host cluster with a {\it Gaia} 
DR2 distance; see Section 3.3 below.

{\it HD 160529:} {\it Gaia} DR2 indicates that this object's new distance is
$2.09^{+0.36}_{-0.27}$ kpc, which is smaller than its usually adopted
value of 2.5 kpc.  Its luminosity moves down by about 30\%, although 
the old value is just outside the 68\% confidence interval.  This places HD~160529 
at the very bottom of the traditional range of luminosities for LBVs.

{\it HD 168607 and HD 168625:} Although HD~168625 is a candidate LBV (section 3.2 below), 
we discuss it here alongside HD~168607.  These two are usually considered as a pair since 
they are only 1$\arcmin$ apart on the sky.  As described in Section~\ref{sec:lit}, 
HD168607 and HD168625 have both been considered members of the same stellar association 
Ser~OB1. HD168607 has long been considered as a confirmed LBV \citep{chentsov80} in the 
group of low-luminosity LBVs \citep{smith04}, sometimes thought to arise from post-RSG 
evolution.  HD168625 is a blue supergiant and considered an LBV candidate based on its
dusty circumstellar nebula, which is observed to have a triple-ring structure that is 
very reminiscent of SN~1987A \citep{smith07}. The new {\it Gaia} DR2 measurements for each 
star are both very consistent with one another ($1.46^{+0.17}_{-014}$ and 
$1.55^{+0.19}_{-0.16}$; supporting their joint membership in Ser OB1) and significantly 
lower than the usual literature value of 2.2 kpc. The new {\it Gaia DR2} distances 
(1.32-1.63 kpc) are consistent (within the uncertainty) with the {\it Gaia} DR1 distance 
of 0.75-1.89 kpc for HD168607 previously reported \citep{ss17}.  With the new distances, 
the luminosity of each star drops by a factor of 2 from literature measurements. 
Importantly, both stars land well below the traditional range of luminosities for LBVs 
and well below the S Dor instability strip.  For both stars, the lower luminosity would 
imply an effective single-star initial mass of around 20 $M_{\odot}$ or below. In the 
case of HD168625, this lower luminosity is very close to that of the progenitor of SN~1987A, 
strengthening comparisons between these two objects that were previously based primarily 
on the triple-ring structure of the nebula around HD~168625 \citep{smith07}.  

{\it P Cyg:} P Cyg has a large excess noise, $\epsilon = 1.1$ mas, and the observed 
parallax is only $\varpi = 0.736$ mas. Therefore, the range of allowed distances is quite 
large; the 68\% HDI goes from 0.98 to 4.25 kpc.  This is consistent with the previous 
estimate of 1.7 kpc.  Unfortunately, we cannot say anything new about the distance and 
luminosity of this LBV due to the large astrometric noise and large DR2 uncertainty.

{\it MWC~930 (V446~Sct):} The distance for this confirmed LBV \citep{miro14}
has a negative parallax in Gaia DR2. A crude way to deal with the 
negative parallax is to treat the parallax uncertainty as a lower limit 
for the distance.   Including both the statistical uncertainty and the excess noise, the 
total uncertainty is $\sigma = 0.24$ mas, and the corresponding lower limit for this 
LBV is 4.2 kpc.  At face value it would seem that the lower limit is farther 
than previously adopted literature value.  However, the more formal statistical 
Bayesian inference gives a distance of $4.5^{+2.4}_{-1.5}$ kpc (Table~\ref{tab:tab1}).  
This value is dominated by the adpted prior.   This is marginally consistent with the 
previous literature value of 3.5 kpc.

{\it G24.73+0.69:} This LBV with a dusty shell has a negative value for the parallax listed 
in {\it Gaia} DR2, and is similarly dominated by the prior.  The statistical uncertainty 
for G24.73+0.69 is only $\sigma_{\varpi} =0.223$ mas, but the excess noise is quite
large, $\epsilon=1.302$ making our formal uncertainty $\sigma = 1.3$
mas.  In this case, the lower limit for G24.73+0.69 is 0.8~kpc. Given this 
large uncertainty, the Bayesian distance in Table~\ref{tab:tab1}, 
$3.0^{+2.6}_{-1.6}$ kpc, is dominated by the prior.  This is lower than the 
literature value of 5.2~kpc, but consistent within the uncertainty, and so this star is 
still marginally consistent with being an LBV on the S Dor instability strip.  The higher 
precision expected in future {\it Gaia} data releases is needed to say more.

{\it WS 1:} WS 1 was classified as a bona-fide LBV by \cite{Kniazev2015} based on 
observations of significant photometric and spectroscopic variability. The {\it Gaia} 
DR2 distance of 2.48$^{+0.73}_{-0.54}$kpc is \emph{significantly} lower than the 
baseline literature value of $\sim$11 kpc, although this object has a negative parallax 
and like the previous two, is dominated by the prior.   However, the literature luminosity 
is also highly uncertain because it is not based on a measurement --- itwas derived by 
{\it assuming} that the star fell on the S Dor instability strip \cite{Gvaramadze2012}, 
so perhaps a large revision is not so 
surprising. {\it Gaia} data suggests that the star is significantly lower in luminosity, 
although again DR3 data are needed for a confident result.

{\it MN48:} MN48 was classified as a bona fide LBV by \cite{Kniazev2016} who identified 
spectroscopic and photometric variability typical of LBVs. There is large excess noise 
in the Gaia DR2 data for this star, and the resulting 68\% HDI spans 1.44 to 5.18 kpc. 
The upper end of this range is marginally consistent with the previous literature value
of $\sim$5 kpc, although again, the literature value was highly uncertain as well.

\subsection{Candidate LBVs}

{\it HD 80077:} This blue hypergiant was included as an LBV candidate by \citet{vg01} 
based mainly on its high luminosity and spectrum.  Its new {\it Gaia} DR2 distance 
is significantly closer than the baseline literature value, reducing its nominal luminosity 
by about 40\%, although its luminosity is still above the S Dor strip.  The {\it Gaia} 
DR2 distance and the original cluster-fitting distance of \cite{Steemers1986} (2.8$\pm$0.4) 
agree within their quoted errors. 
 
{\it SBW1:} As in the cases of Sher~25 and HD~168625, this blue supergiant is an LBV 
candidate because of its circumstellar nebula that bears a remarkable resemblance to the 
ring nebula around SN~1987A.  In fact, it has been argued that SBW1 is in some respects 
the best Galactic analog to the progenitor of SN~1987A, in terms of both its nebula and 
the properties of the weak-winded central star \citep{smith13,smith17}.  Based on various 
considerations, \citet{smith17} favored a distance of about 7 ($\pm$1) kpc for SBW1.  
The {\it Gaia} DR2 parallax indicates a distance of 5.2$^{+1.4}_{-1.0}$ kpc 
(Table~\ref{tab:tab1}).  This distance is closer, but the range of 4.0-6.8 kpc overlaps 
with the uncertainty of 7 ($\pm$1) kpc for the previous distance.  This 
new distance corresponds to a somewhat lower effective single-star initial mass of 13-15 
$M_{\odot}$, as compared to $\sim$18 $M_{\odot}$ for the progenitor of SN~1987A
\citep{arnett89}.  This confirms the notion that SBW1 is an analog of the progenitor of 
SN~1987A and its nebula \citep{smith13}, although at somewhat lower initial mass.  
Importantly, the new {\it Gaia} DR2 distance and uncertainty confirm that although SBW1 is 
seen to be projected amid the young Carina Nebula at around 2 kpc, it is in fact a luminous 
background blue supergiant star and not a member of the Carina nebula population.

{\it Hen 3-519:} The large {\it Gaia} DR2 distance for Hen~3-519 contradicts
the closer distance from {\it Gaia} DR1 reported by \citet{ss17}, which was
around 2 kpc.  The new {\it Gaia} DR2 distance is $7.6^{+2.5}_{-1.7}$ kpc, and
is consistent with the traditionally assumed value of around 8 kpc.  This indicates 
that Hen 3-519 is still a very luminous LBV candidate.  Its new value barely overlaps 
with the S~Dor instability strip within the uncertainty, although it may be below the S~Dor
strip (especially for some proposed locations of the S DOr strip, as shown in 
Figure~\ref{fig:hrd}.  The reason why the DR1 distance was too close likely has to do 
with the interpretation of the large uncertainty.  The uncertainty in 
parallax for Hen~3-519 was large to begin with, having a DR1 value of $\varpi$ = 
0.796 $\pm$0.575 mas.  Adding a correction of 0.3 mas to this 
uncertainty, as noted by \citet{ss17}, would give a negative parallax.  
In hindsight, taking the uncertainty as indicating a lower limit to the 
distance would have been a better choice. \citet{ss17} adopted the prior 
distribution for stars in the Milky Way from \citet{abj16}, which was 
intended to account for systematic underestimate of the luminosity, but 
that prior distribution may have been inappropriate for a distant LBV.

{\it Sher~25:} The blue supergiant Sher 25 is of interest because of
its ring nebula that resembles the equatorial ring around SN~1987A,
and because of its projected proximity to the massive young cluster
NGC~3603.  The large distance of 6.6 kpc (Table~\ref{tab:tab1}) is remarkably 
consistent with previously assumed values.
This is important because Sher 25 was thought to be
considerably more luminous than the progenitor of SN~1987A (also shown
in Figure~\ref{fig:hrd}), with a luminosity that corresponded to about
twice the initial mass ($\sim$40 $M_{\odot}$ vs. 18 $M_{\odot}$).  It appears to be 
below or barely overlapping with the S~Dor instability strip.  These results are 
suggestive that Sher~25 is too massive to be considered as a good analog of SN~1987A's 
progenitor in terms of initial mass.  Importantly, at this luminosity,
the ring around Sher~25 has probably not arisen from a fast blue
supergiant wind that swept into a previous red supergiant (RSG) wind,
since in that scenario the surrounding nebula would probably be more massive and younger.
This is in agreement with its chemical abundances, which are
inconsistent with the level of enrichment expected if it had passed through a
previous RSG phase \citep{smartt02}.

{\it $\zeta^1$ Sco:} This is a blue hypergiant star that was included as an ``ex-dormant" 
LBV (i.e. an LBV candidate) by \citet{vg01} because of its hypergiant-like spectrum and 
microvariability. It was previously assumed to be at a distance of around 2 kpc, placing 
its luminosity just above the S~Dor instability strip.  The excess noise for this star is 
quite large, $\epsilon=0.95$ mas, making the distance uncertainty quite large and dominated 
by the prior.  The old value for the distance and the Gaia DR2 most likely distance are
above the S~Dor instability strip.  However, the new uncertainty
overlaps the instability strip.

{\it HD 326823:} This blue supergiant was considered as another
``ex-dormant'' LBV (candidate) by \citet{vg01}, again because of its spectrum 
and microvariability.  More recently, it has been suggested to 
be a close binary system with a period of 6.1~d \citep{richardson11}. 
Its new {\it Gaia} DR2 distance
is reduced from the old value by about 37\%, reducing its luminosity
to less than half its previous value.  The error bar on the {\it Gaia} DR2
distance (1.18-1.38 kpc) is small enough that this is a significant revision.  Its
corresponding effective single-star initial mass is reduced from
$\gtrsim$25 $M_{\odot}$ to 17-18 $M_{\odot}$.  This is yet
another case of an LBV candidate with stellar properties very similar to
those of SN~1987A's progenitor or the putative surviving companion of
SN~1993J.  This similarity combined with its status as a close binary make this a 
potentially very interesting target for comparison with the progenitor systems of 
SN~1987A and SN~1993J.

{\it Wray 17-96 (B61):} This classic B[e] supergiant is considered to be an LBV candidate 
based on its dusty circumstellar shell nebula discovered by IR surveys \citep{egan02}.  
Its previously adopted distance of 4.5 kpc would imply an extreme luminosity for this object
above 10$^6$ $L_{\odot}$, placing it in the regime of classical high-luminosity LBVs, 
although it has not exhibited LBV-like variability.  Unfortunately, the large excess 
astrometric noise limits the {\it Gaia} DR2 distance between 1.2 and 7.2 kpc.

{\it HDE 316285:} This star is sometimes considered an LBV candidate due
to is remarkable spectral similarity to $\eta$~Car \citep{hillier01}
and its dusty nebula \citep{clark05,morris08}. Although the uncertainty in
distance is large, the new {\it Gaia} DR2 parallax suggests that HDE~316285
is significantly farther away than previously assumed, moving it from about 1.9 kpc
out to about 5 kpc.  This raises its luminosity by more than a factor
of 6.  As such, it is pushed well above the upper luminosity limit for
RSGs and into the regime of the classical LBVs, becoming potentially
even more luminous than AG~Car (although, again, the error bar is
large).  This larger distance and higher luminosity may help explain
why the spectrum of HDE~316285 has such an uncanny resemblance to
$\eta$~Car \citep{hillier01}. Interestingly, \citet{morris08} speculated 
that HDE~316285 may be coincident with Sgr D near the Galactic center at 
a distance of around 8 kpc.  This is farther than the most likely {\it Gaia} 
DR2 distance, but permitted within the uncertainty, although it is unclear if 
its line of sight extinction is consistent with a distance this large.

{\it HD~168625:} see above (Section 3.1).

{\it AS 314:} This star was considered to be an LBV candidate based on
its presumed high luminosity at a large 8-10 kpc distance, its
hypergiant-like spectrum, and dust excess \citep{miro00,vg01,clark05}.
The new {\it Gaia} DR2 parallax moves its distance from 8 kpc to only about
$1.5^{+0.16}_{-0.13}$ kpc, and with no excess astrometric noise, lowering 
its luminosity by a factor of 25 to only log($L$/$L_{\odot}$)=3.5.  Its new 
luminosity is so low that it cannot  be plotted in Figure~\ref{fig:hrd} because 
it is off the bottom of the plot (even its upper error bar is below the bottom 
of the plot).  This is probably a post-AGB object from an intermediate-mass star, 
and it is most likely not related to LBVs.

{\it MWC~314:} This is one of two sample stars with both a relatively small uncertainty on 
the distance, and where the distance has {\it increased} compared to values adopted in the 
literature.  This LBV candidate has moved from about 3 kpc to $4.0^{+1.03}_{-0.7}$ kpc 
(Table~\ref{tab:tab1}), roughly doubling its luminosity.  This moves it about one sigma off 
the S~Dor instability strip and makes it similar to $\eta$~Carinae on the HR diagram 
(Figure~\ref{fig:hrd}).  The distance calculated from the parallax makes it seem quite 
likely that MWC~314 may be one of the most luminous stars in the Milky Way.

{\it W51 LS1:} This luminous blue supergiant was added to the list of LBV candidates by 
\citet{clark05}, based on its supergiant spectrum and high luminosity.  It has shown no 
variability or obvious shell nebula, but does have a near-IR excess.  As described in 
Section~\ref{sec:lit}, the distance to the W51 complex has a complex history.  
The new {\it Gaia} DR2 parallax indicates a smaller distance for this source than 
most previous measurements: $2.5^{+2.4}_{-1.3}$.  The wide HDI is due to the large 
excess astrometric noise.  With such a large excess noise, the posterior distribution 
is heavily influenced by the prior distribution, and while inconsistent with our 
baseline literature value $\sim$ 6 kpc, does just barely overlap with recent maser proper 
motion measurements of \cite{Xu2009,Sato2010}. Intriguingly, our {\it Gaia} measurements 
also show significant overlap with the spectroscopic parallax measurement of 
\cite{Figueredo2008}.  W51 LS1 revised position lies within the region of normal
blue supergiants and B[e] supergiants (green oval) in the HR diagram, but the large 
uncertainty still encompasses the lower end of the S~Dor instability strip 
(Figure~\ref{fig:hrd}).

{\it G79.29+0.46:} This is an LBV candidate with a dust shell.
The Gaia DR2 distance $1.9^{+1.4}_{-0.8}$ kpc (Table~\ref{tab:tab1}) is consistent 
with the literature distance of 2 kpc.  The distances are consistent, and so G79.29+0.46 
is still likely a massive star, and its dust shell is not a planetary nebula.

{\it Cyg OB2 \#12:} Cygnus OB2 \#12 is a B hypergiant that is usually
considered as an LBV candidate because of its extremely high
luminosity and cool temperature \citep{clark05,massey01,hd94}.  It has
rather mild variability, so \citet{vg01} classified it in the group of
``weak-active'' S~Doradus variables (meaning low-amplitude $<$0.5 mag
variability).  Its status as one of the most luminous stars in the
Milky Way (e.g., \citealt{dejager98}) is based on its presumed
association with Cyg~OB2 at about 1.7 kpc \citep{clark05,clark12}. Recently, however, \citet{berlanas19} have questioned this association.  \citet{berlanas19} find two likely populations along the same line of sight from a recent analysis of {\it Gaia} DR2 data, with one at 1.76 kpc (close to the traditional distance for Cyg OB2) and a foreground group at 1.35 kpc.  
The excess astrometric noise for this star is quite large, 0.588 mas.  The 
measured parallax is $\varpi$=1.178 ($\pm$0.128) mas, so while the measured 
parallax is larger than the excess noise, the excess noise does significantly impact 
the uncertainty.  Moreover, the RUWE value of 1.52 makes this the only source in our sample that is bigger than the value of 1.4, above which \citet{lindegren2018} caution against. The HDI for Cyg OB2 \#12 is between 0.6 and 2.2 kpc.  Hence, {\it Gaia} DR2 is consistent with the literature distance, or the two diufferent distance of subgeroupos along the same line of sight of 1.76 or 1.35 kpc \citep{berlanas19}.

\begin{figure}
\includegraphics[width=3.1in]{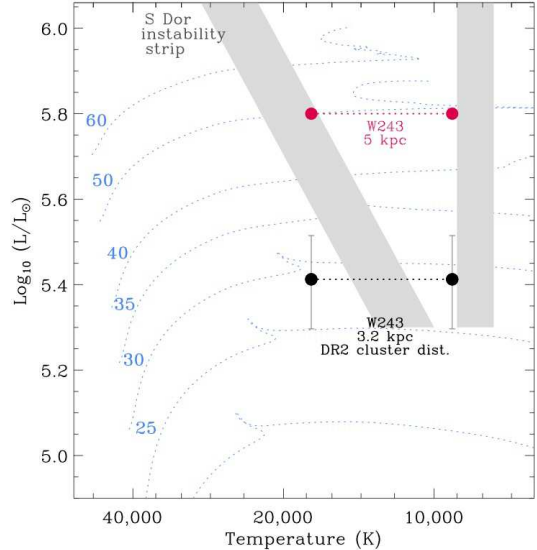}
\caption{An HR diagram similar to Figure~\ref{fig:hrd}, but showing only W243 based on the old distance of around 5 kpc (red) and using a distance of 3.2 $\pm$0.4 kpc (black) based on the distance inferred for the whole Wd1 cluster from DR2 data \citep{mojgan18}.  The reference single-star models are the same as in Figure~\ref{fig:hrd}.}
\label{fig:wd1}
\end{figure}

\subsection{Westerlund 1 and W243}

W243 is a confirmed LBV that is thought to be associated with the young 
massive star cluster Westerlund 1 \citep{cn04,clark05}. It exhibits photometric 
variability and changes in spectral type that have been interpreted as LBV-like 
variability \citep{cn04}, but the variability that has been observed is not 
conclusively due to S~Dor-like cycles.  With a often-presumed distance of 5 kpc 
for that cluster, W243 would be a luminous LBV with an effective single-star 
initial mass around 50 $M_{\odot}$, similar to S~Dor.  With that presumed mass, 
it helps define the inferred turn-off mass, young age, and high total stellar 
mass of Wd1.  In {\it Gaia} DR2, W243 has a large astrometric excess noise, 
$0.582$ mas, which is almost as large as the measured parallax, 
$\varpi = 0.979 \pm 0.165$ mas.  As a result, the {\it Gaia} DR2 HDI for W243 
alone is unfortunately large, between 0.83 and 4.15 kpc, the upper end of which 
is closer than the typical literature value (although to be fair, previous literature 
estimates had a wide dispersion from 2-5 kpc as well, a dispersion that is often not quoted).  

On the other hand, if W243 is truly a member of the Wd1 cluster, then the numerous 
other members of the cluster (many of which do not have such large astrometric noise) 
can be used to reduce the uncertainty in distance.  A full analysis of the DR2 
results for a sample of stars in Wd1 is beyond the scope if this paper, but we have 
conducted this analysis and discuss it in a separate article \citep{mojgan18}.  
In that study, we find a well-determined {\it Gaia} DR2 distance to 
Wd1 of 3.2 $\pm$0.4 kpc.  This cluster-based distance would be a less extreme 
revision to its distance and luminosity than for the {\it Gaia} DR2 data for the 
star W243 alone, but also clearly reduced from the usually adopted value around 5 kpc.  
At the new cluster distance, W243 would have a lower luminosity, appropriate to an 
evolved single star of around 25-30 M$_{\odot}$ (see Figure~\ref{fig:wd1}) instead 
of 50 $M_{\odot}$ (the implied cluster turn-off mass would be lower as well, and the 
cluster age would be older than previously assumed so that it is not such a young and 
massive cluster after all; see \citealt{mojgan18}).   W243 would be near the bottom of 
the traditional range of luminosities for LBVs, close to where the S Dor strip meets 
the constant-temperature outburst strip.  However, at this low luminosity, W243's 
presumed hot temperature around 17-18 kK indicated by its B2 spectral type in 
quiescence \citep{cn04,clark05} would move it off the S~Dor instability strip at 
quiescence, and its observed variation in temperature would be too large to be 
consistent with traditional expectations for LBVs.   For now, 3.2 ($\pm$0.4) kpc 
is our preferred distance to W243 because it is more precise, although formally, 
it is fully consistent with the individual DR2 value we find between 0.83 and 
4.15 kpc.  Although{\it Gaia}  DR2 data are not consistent with a large distance 
of 4.5 or 5 kpc that is often adopted in the literature, the new DR value is 
consistent with the original claimed distance of 2 kpc $<$ d $<$ 5.5 kpc \citep{clark05}.

The case of W243 illustrates the utility in having an independent estimate of the distance based on the parallax for many associated cluster members, especially in individual cases where high $\epsilon$ prohibits a reliable individual parallax.  Unfortunately, however, it appears that many LBVs simply do not reside in obvious clusters \cite{st15}. There are some LBVs with possible host clusters that have been noted, such as Wra 751.  Even though Wra~751 appears to be overluminous for this cluster, it may help to more tightly constrain the distance and true age and initial mass of this object.  Another possible case to investigate is $\zeta^1$Sco.  HD~168625 and HD~168607 may be associated with Ser OB1, as noted earlier, but these already have quite good {\it Gaia} DR2 distances (with $\epsilon$=0.0).

\section{DISCUSSION}

About half of the LBVs and LBV candidates included in {\it Gaia} DR2
have literature distances that are within the 68\% HDI.
These are the LBVs HR~Car, Wra~751, 
P~Cygni, MWC930, MN48, and G24.73+0.69, and the LBV candidates Hen~3-519,
Sher~25, SBW1, $\zeta^1$~Sco, WRAY~17-96, G79.29+0.46, and CYG~OB2~\#12.  While the uncertainty overlaps with literature estimates, the value for the distance from {\it Gaia} DR2 is lower in most of these cases.
%%%%%%%%%%%%%%%%%%%   
%Actual LBVs
%Within HDI
% MN48
%HR Car, 
%WRA 751
%P Cyg
%MWC 930
%G24.73
%All but HR Car have large excess noise.
%%%%%%%%%%%%%%%%
%Closer
% WS1
%AG Car
%W243
%HD 160529
%HD 168607
%All but W243 and WS1 have no excess noise.
%
%
%%%%%%%%%%%%%%%%
%Candidates
%Within HDI
%Hen 3-519
%Sher 25
%zeta sco
%WRAY 17-96
%G79
%Cyg OB2
%%%%%%%%%%%%%%%
%Closer
%HD80077
%SBW1
%HD 326823
%HD 168625
%AS 314
%W51 LS1
%%%%%%%%%%%%%%%
%Farther
%HD 316285
%MWC 314

Two LBV candidates have had their distances and luminosities increase significantly 
as a result of DR2.  MWC314 has a significantly increased distance that makes its 
luminosity comparable to that of $\eta$~Carinae.  HD~316285 also has a significantly 
increased distance, raising its luminosity above 10$^6$ L$_{\odot}$.  

For almost all the objects where the {\it Gaia} DR2 distance is significantly revised, 
however, we find that the objects have moved closer and their luminosity is lower than 
traditionally assumed.  This reduction applies to 10 objects, including both confirmed and
candidate LBVs.  Of the five confirmed LBVs that have moved significantly closer (AG~Car, 
W243, HD~160529, WS1, and HD~168607), only W243 and WS1 have large excess noise.  As we noted, 
however, if we adopt the new {\it Gaia} DR2 parallax for the host cluster Wd1 as the 
distance to W243 \citep{mojgan18}, then this object is also confidently closer and less 
luminous, but with a much smaller uncertainty.  Five of the LBV candidates are moved to a 
significantly closer distance and lower luminosity, although one of them (AS314) has such 
a small distance and low luminosity that it is probably not related to LBVs).  
Overall, with a few getting brighter, many not changing significantly, and $\sim$10 
shifting to significantly closer distances and lower luminosities, 
the net effect is a widening and an overall downward shift of the observed luminosity range 
for Galactic LBVs.  This broad conclusion is true whether or not we include sources with 
excess astrometric noise.  Obviously the situation is expected to improve with DR3, but that is 
several years in the future, and it is worthwhile to consider the implications for LBVs now.

\subsection{Luminosity range and the S Dor strip}

With a larger spread in LBV luminosity range that now extends to lower 
luminosity than previously assumed,
there are two divergent ways to interpret the result.  One option is
simply that these lower-luminosity stars were mistakenly classified as
LBVs.  The other view is that the original definition of the S~Doradus
strip, based on only a few objects, might have been too narrow; it might
therefore fail to capture the diverse range of real physical
variability and mass loss exhibited by luminous, blue, and irregularly
variable stars.  Which of these two options is chosen might have
important implications for understanding the range of initial masses
that yield LBVs and the physical mechanism(s) governing their
instability and mass loss.  Both options have some subjectivity.

Following the first option, we might decide to strip these
lower-luminosity stars of their LBV or LBV candidate status, demoting
them to ``normal'' blue supergiants or B[e] supergiants and thus
preserving the S~Doradus instability strip.  This
demotion may be valid for some objects, where the main
motivation for including them as LBV candidates in the first place was
their high luminosity (like AS~314).  However, it is less
appealing to simply discount the lower-luminosity candidates with
dusty shell nebulae, because these nebulae indicate substantial
episodes of previous mass loss that are relatively rare among blue
supergiants.  It is also not so easy to discount lower luminosity
stars that have strong emission-line spectra that resemble their very
luminous cousins, since these emission-line spectra indicate
strong current mass loss.  Moreover, it is not so easy to ``unconfirm''
Wra~751 or the lower-luminosity objects that have been confirmed
as LBVs based on their variability (HD~168607, W243, and WS1).  If these are 
not lower-luminosity LBVs, then what type of blue irregular variable are they?

We note that previous efforts to classify a star as an ``LBV'', ``LBV
candidate'', or ``neither'' (not to mention various subtypes like
classical S~Doradus stars, SN impostors, ex-dormant, weak active,
P~Cygni stars, etc.) have been somewhat arbitrary and inconsistent
among authors \citep{conti84,hd94,vg01,smith04,clark05}.  This
reflects the fact that LBVs are rare stars and that each one has
some unique peculiarities.  To classify them in a group or subgroups
requires one to make choices about which observed properties to
emphasize in a definition.  The original definition of an LBV by
\citet{conti84} was a hodgepodge of many different types of massive
and variable hot stars -- basically ``not Wolf-Rayet stars'' and ``not
red supergiants'' (in fact, Conti used the term ``other'') -- including the
Hubble-Sandage variables, S Doradus variables, $\eta$ Car-like
variables, P Cygni stars, etc.  The motivation was that all these
stars may play a similar transitional role in evolution once O-type
stars leave the main sequence, and it is potentially useful to discuss
them together.  It is also a convenient oversimplification for the purpose of discussion. 
As study of these stars intensified, some
observers, guided by stellar evolution models for single stars, 
favored a more precise definition of what is an LBV so that
only very few objects were included \citep{wolf89,hd94}, whereas others
chose to proliferate LBV subtypes to accommodate some of the diversity
in observed characteristics \citep{vg01}.  Some objects were included
as LBVs or LBV candidates based on much more limited information than
for the classical LBVs, as in the cases of the Galactic Center sources.  In light of the fact
that we still do not understand the physical mechanism that drives LBV
variability or their place in evolution, it may be wise to lean toward
being inclusive with respect to this diversity.  Regardless of the name we choose to give them, revised
distances and luminosities from {\it Gaia} DR2 seem to indicate that
blue supergiants at lower luminosity than previously thought can also
suffer episodes of mass ejection, variability, instability, and strong winds that
could be similar to traditional expectations for LBVs.

A few objects are also found to be off the S~Dor instability strip,
but {\it above} it.  These include $\eta$ Car (this has been known for
a long time), MWC~314, HDE~316285, and HD~80077.  The Pistol Star and FMM~362 
were also known to be well above the S~Dor instability strip, similar to $\eta$~Car.  
Demoting the lower-luminosity stars from the class of
LBVs would not change the fact that these more luminous stars are also
off the S~Doradus instability strip, again arguing that its definition
may have been too narrow in the past.

Whether or not revised luminosities land a star on the S~Dor
instability strip depends, of course, on exactly where we choose to
put that instability strip. Since the defining S~Doradus variables,
AG~Car and HR~Car, have slightly revised distances, perhaps the
location of the S~Dor strip needs to be adjusted as compared to the
original position \citep{wolf89}.  As noted earlier, \citet{groh09b}
presented a revised S~Dor strip defined by detailed modeling of
physical parameters for AG~Car and HR~Car in their hot quiescent
states (which we have slightly adjusted in Figure~\ref{fig:hrd} based
on their new {\it Gaia} DR2 distances).  If this placement of the S~Dor
instability strip is adopted, then $\eta$~Car and MWC~314 fit nicely
along an extension of its slope, as do P~Cygni and HD~168625.  This steeper slope also
encompasses the general locations of B[e] supergiants and many LBV
candidates.  However, with that steeper slope, many other LBVs are
then left far off the S~Dor strip, including most of the known
extragalactic LBVs, as well as W243, HD~160529, HD~168607, and
G24.73+0.69.  Wra~751 is far off the S~Dor strip no matter what.

It seems difficult to escape the conclusion that the S~Dor instability 
strip must be broader and must extend over a wider luminosity range 
than previously appreciated.  How shall we interpret this?  One possible 
option is that the LBV instability does indeed occupy a wider spread of
luminosity and $T_{\rm eff}$ than the narrow strip originally defined by 
\citet{wolf89} or the revised version proposed by \citet{groh09b}.  When 
one examines Figure~\ref{fig:simple}, where no extragalactic LBVs are 
plotted and where we show no S~Dor strip to guide the eye, it is not 
obvious that LBVs reside along any strip at all. 

Figures~\ref{fig:simple}, \ref{fig:oldnew}, and \ref{fig:hrd} give the impression that 
the zone of instability for LBVs and related objects might include everything redward 
of the terminal age the main sequence, over to about 8000~K, spreading both above
and below the S~Dor strip (objects that are luminous and variable but
cooler than 8000~K are not called LBVs because they are yellow or
red).  In other words, one might simply extend the locus of normal
BSGs and B[e] supergiants upward to include the classical LBVs as
well.  This zone encompasses LBVs and LBV candidates, but also
includes many blue supergiants that are not highly variable and do not
have significant circumstellar material.  In this view, perhaps the classical 
LBVs are just the most extreme end of a continuum of diverse variability and mass loss.

Whether or not a star in this zone is an LBV may depend on its evolutionary 
history, as well as our choice for the threshold of variability needed to call 
that star an LBV.  Some LBVs could conceivably be on a
post-RSG blue loop, where previous strong mass loss as a RSG has increased
their L/M ratio, making them more unstable.  This may work for the
stars in the 30-40 $M_{\odot}$ initial mass range
\citep{hd94,smith04}, but at lower luminosities, stars remain far from
the classical Eddington limit.  The other viable option is that some
stars arrive in this zone through single-star evolution whereas others
arrive there as a product of binary interaction, or that they have different 
rotation rates.  Mass accretion and
spin up through binary mass transfer or stellar mergers may provide a
means for only some selected stars in this zone to experience peculiar
and episodic mass loss, anomolous enrichment, rapid rotation, and
instability \citep{kg85,jsg89,justham14,st15,smith16,mojgan17}.  There
is no clear reason why such effects would be limited to a narrow zone
coincident with the S Dor instability strip, so in a binary context, the
wider spread of luminosity would not be surprising.

\subsection{Lower luminosity LBVs}

{\it Gaia} DR2 reveals a handful of LBVs and candidate LBVs that reside at lower luminosities than previously realized, well below log($L$/$L_{\odot}$)=5.3 where the nominal S Doradus instability strip joins the constant temperature outburst temperature of LBVs.    Why have analogs of these lower-luminosity LBVs not been found in the LMC/SMC or M31/M33 (the purple
sources in Figure~\ref{fig:hrd})?  This might easily be a selection
effect since it is harder to detect subtle variability in fainter
stars, especially if one is interested in the most luminous stars.
Alternatively, ``LBV-or-not'' classifications in these galaxies may
have been biased to high luminosities (the brightest objects are
deemed LBVs, while fainter blue stars with variability may have been
ignored or called something else).  There may also be a real physical
effect: perhaps whatever mechanism is
responsible for the LBV instability (such as Fe opacity; e.g., 
\citealt{grafener12,jiang18}) can be triggered
at different luminosities in a higher-metallicity environment or at 
different rotation rates.  Deciding between these options is difficult, 
and a renewed and unbiased effort to characterize variable stars in these 
nearby galaxies may be warranted.  This is an area where the {\it Large 
Synoptic Survey Telescope} may provide a significant advance.  We note that
recently, such a study has been undertaken for M51 by
\citet{conroy18}. Using multi-epoch {\it HST} data, they found a
continuum of variability over a wide luminosity range for luminous
stars, where the observed diversity of variability among luminous blue
stars was considerably broader than the narrow definition of traditional 
S Dor variables.

The possible existence of LBV-like instability at lower initial mass
and lower luminosity than previously recognized has at least three
broader implications.

1.  {\it Physical cause of LBV instability:} The traditional interpretation 
for the cause of normal S~Dor instability has been that these stars are 
unstable because of their proximity to an opacity-modified Eddington limit 
\citep{lf88,uf98}. Single-star models suggest that a star of $\sim$60 
$M_{\odot}$, for example, will develop a progressively more LBV-like 
spectrum and instability as it approaches the Eddington limit in its mass-loss evolution 
\citep{groh14,jiang18}. This may work for the most luminous LBVs, but it may not 
work so well for the lower-luminosity examples in the 30-40 $M_{\odot}$ 
initial mass range.  As noted above, these might plausibly reach a similar 
instability if one invokes rather severe previous RSG mass loss, so that they are 
now in a post-RSG phase \citep{hd94,smith04,groh13,groh14}.  {\it Gaia} DR2 
distances now suggest that there are LBV-like stars at even lower luminosities 
(in the 10-30 $M_{\odot}$ initial mass range).  At such low luminosities, this
near-Eddington instability doesn't work because their previous RSG
mass loss is not strong enough, and their luminosities are not high
enough \citep{bd18}.  We must either conclude that they have a separate 
instability mechanism unrelated to the Eddington limit, or that perhaps some 
other instability governs all the LBVs.  As noted above, post-merger or
post-mass transfer evolution may populate the whole relevant range of
luminosities with massive blue stragglers.  This may be an important clue.

2.  {\it Relation to low-luminosity SN impostor progenitors like SN
  2008S:} The possible existence of LBVs that push to lower
luminosities and initial masses than previously thought may have
interesting implications for a subset of SN impostors similar to the
well-studied object SN~2008S.  SN impostors were generally thought to
be related to giant eruptions of LBVs
\citep{hds99,vd05,vd06,vdm12,smith11}.  However, a few transients in
the past decade, highlighted by the prototypes SN~2008S and NGC~300-OT
\citep{prieto08,prieto+08,bond09}, had dust-obscured progenitors.  These had
lower inferred luminosities and lower initial masses than traditional LBVs, so 
these have been suggested to be transients that arise from super-AGB
stars with initial masses around 8-10 $M_{\odot}$, including electron
capture SNe \citep{thompson09,boticella09}.  However, if LBVs actually extend to lower 
luminosity, then they might also be dust-enshrouded
LBV-like supergiants that reside at somewhat lower luminosity than
previously recognized \citep{smith09,berger09,bond09}.  For
NGC~300-OT, the surrounding stellar population points to an age
appropriate to an initial mass of 12-25~$M_{\odot}$
\citep{gogarten09}, inconsistent with an electron capture SN (8-10
$M_{\odot}$) or a transient associated with an even lower-mass star.
This would, however, be consistent with the implied initial masses for
the lower-luminosity Galactic LBVs like W243, and HD~168607 found here.
Perhaps these lower-luminosity Galactic LBVs are likely progenitors
for some of these SN~2008S-like events, or products of them if they
are merger events.  The spectra for many of these objects look quite
similar at various points in their evolution, including objects that
have been suggested to be stellar mergers \citep{smith11,smith+16}.  
On the other hand, the SN impostors may be a mixed-bag across a wide 
mass range, since some, like SN~2008S itself and SN~2002bu, have surrounding 
star formation histories that translate to ages appropriate for initial masses 
less than 8 $M_{\odot}$ \citep{williams18}.

3.  {\it SN progenitors with pre-SN mass loss:} The group of
lower-luminosity Galactic LBVs and LBV candidates have interesting
potential implications for some types of SN progenitors.  First, we
have noted that several LBV-like stars seem to be quite close
to the location of SN~1987A's progenitor on the HR diagram, and few of
these even have similar ring nebulae.  This adds to speculation that
some sort of LBV-like instability and mass loss could have played a
role in forming the nebula around SN~1987A \citep{smith07}.
Previously, it was thought that the lower bound of LBV luminosities
did not extend low enough to include SN~1987A, but {\it Gaia} DR2 
shows that it reaches even lower.
Second, there has been much discussion about LBVs as possible
progenitors of Type IIn supernovae (SNe IIn), because their dense
circumstellar material (CSM) seems to require some sort of eruptive
pre-SN mass loss akin to LBV eruptions (see review by
\citealt{smith14} and references therein).  In seeming contradiction,
host galaxy environments surrounding SNe~IIn (and also SN impostors)
do not favor very high mass stars in very young regions
\citep{aj08,anderson12,habergham14} 

Even the special case of
SN~2009ip, with a very luminous and eruptive LBV-like progenitor, is
out in the middle of nowhere, with no sign of recent star formation
\citep{smith+16b}.  If the LBV phenomenon extends to much lower masses
than previously thought, then perhaps SNe IIn can arise from LBV-like
progenitors over a wide range of initial masses that even overlaps
with progenitors of normal SNe II-P.  The lower-luminosity LBVs in
Figure~\ref{fig:hrd} overlap with single-star evolutionary tracks as
low as 10-20 $M_{\odot}$.  If they are the results of mass gainers or
mergers in binary systems, then their true initial masses may extend
even lower, and their lifetimes may potentially be quite long 
\citep{st15,smith16,mojgan17}.  Such
LBV-like progenitors of SNe~IIn originating from this lower-mass range
might vastly outlive any main-sequence O-type stars that could ionize
surrounding gas, possibly explaining the lack of correlation between
SN~IIn locations and H$\alpha$ emission
\citep{aj08,anderson12,habergham14}.

This last point seems to be in general agreement with the relative
isolation of LBVs on the sky as compared to O-type stars \citep{st15}.
Moreover, some of the firmer distance estimates for Galactic LBVs
reported here have implications for the isolation of LBVs and
implications for their evolutionary origin in binary systems
\citep{st15,smith16,mojgan17}.  
%W243 as among the very few cases of LBVs 
%located in a massive young star cluster.  Its {\it Gaia} DR2 distance is 
%vastly lower than for the previously adopted distance to the cluster, suggesting 
%that it is not actually a cluster member, but a line-of-sight coincidence.  The 
%closer distance makes W243 far less luminous (and less massive), and therefore 
%more likely (by number) to be seen along the line of sight to a background cluster.
%Although W243 was assumed to be a member of a young massive cluster, it is 
%unlikely that dense source crowding has caused an anomalous parallax measurement. 
%W243 is not seen to be located near the crowded center of Wd1.  It is seen to be 
%projected well outside the cluster core where the density of sources is much lower.  
%FMM~362 is about 1{\arcmin} (2.5 pc at 8.5 kpc) to the NE of the Quintuplet's center in Figure 1 of \citet{fmm}, and 
%W243 is about 2{\arcmin} (2.6 pc at 4.5 kpc) to the SE of Wd1 Figure 1 of 
%\citet{clark05b}.  This offset may make it more believable that W243's association 
%with its cluster on the sky could be due to a chance a projection rather than physical
%membership.  As noted above, Wd1 is seen down a tangent point in the Carina arm.

\subsection{LBVs and clusters}

Among confirmed unobscured LBVs that are found in young massive clusters, now only $\eta$~Car 
remains as a confident association, although its luminosity is so high that it is 
consistent with being a blue straggler as compared to the surrounding stars in 
Tr~16. The earliest-type main sequence star in Tr16 is O3.5~V with an implied initial 
mass of around 60 $M_{\odot}$, whereas $\eta$ Car has an equivalent single-star initial 
mass of around 200 $M_{\odot}$ or more.  Indeed, there are a number of other clues, such as 
light echoes from the Great Eruption, that $\eta$ Car's present state is inconsistent with single-star 
evolution and might instead be the result of a stellar merger event \citep{smith18a,smith18b}. 
Although W243 is probably in the cluster Wd1, 
the revised nearer distance to Wd1 from {\it Gaia} DR2 \citep{mojgan18} means that the 
cluster is not as young and massive as previously thought.  Wra~751's distance makes it 
marginally consistent with its presumed host cluster \citep{pasquali06}, but this is not 
a young massive cluster either, and this is actually a
problem for the single-star scenario.  This is because Wra~751 has a
luminosity indicating an effective single-star initial mass that is
more than 2 times higher than the turnoff mass inferred from the late
O-type stars still on the main sequence in that same cluster.   AG~Car and Hen~3-519 are at a 
large distance, but this means that they are not at the same
distance as O-type stars that appear near them on the sky in the Car
OB association, which is at around 2 kpc \citep{ss17}.  This makes
their apparent isolation even worse, and the discrepancy is
exacerbated by the fact that the higher distance also gives them a
higher luminosity and shorter lifetime.  It is remarkable that AG~Car
has a luminosity consistent with an initial mass of around 80
$M_{\odot}$, but it is not known to be associated with any O-type stars
at a similar distance.  More detailed investigations of any possible
birth populations associated with LBVs could be illuminating.  

Of course, there are a few luminous LBVs and candidate LBVs known in highly obscured 
massive clusters, such as the clusters in the Galactic Center.  However, their significance 
is difficult to judge, and their implications for LBV evolution are unclear, since these 
were discovered as LBV-like stars based on studies of the clusters and are therefore a 
heavily biased sample. As noted by \citet{st15}, there is no available census of 
visually-obscured LBVs in the field or a census of highly obscured O-star 
populations around them that can be used to infer what fraction obscured LBVs 
in the Galactic plane avoid clusters.

\section*{Acknowledgements}

\scriptsize

We thank an anonymous referee for a careful reading of the manuscript and helpful comments.
Support for NS was provided by NSF award %AST-1312221 and 
AST-1515559,
and by the National Aeronautics and Space Administration (NASA)
through HST grant AR-14316 from the Space Telescope Science Institute,
which is operated by AURA, Inc., under NASA contract NAS5-26555.
%This research has made use of the SIMBAD data base, operated at CDS,
%Strasbourg, France.
Support for MA and JWM was provided by the National Science Foundation under 
Grant No. 1313036.  Support for this work was provided to MRD by NASA through 
Hubble Fellowship grant NSG-HF2-51373 awarded by the Space Telescope Science 
Institute, which is operated by the Association of Universities for Research in 
Astronomy, Inc., for NASA, under contract NAS5-26555. MRD acknowledges support 
from the Dunlap Institute at the University of Toronto.
JHG acknowledges support from the Irish Research Council New Foundations Award 
206086.14414 ``Physics of Supernovae and Stars".

This work has made use of data from the European Space Agency (ESA) mission
{\it Gaia} (\url{https://www.cosmos.esa.int/gaia}), processed by the {\it Gaia}
Data Processing and Analysis Consortium (DPAC,
\url{https://www.cosmos.esa.int/web/gaia/dpac/consortium}). Funding for the DPAC
has been provided by national institutions, in particular the institutions
participating in the {\it Gaia} Multilateral Agreement.

\end{document}